\documentclass[]{spie}  %>>> use for US letter paper
\usepackage{amsmath,amsfonts,amssymb}
\usepackage{graphicx}
\usepackage[colorlinks=true, allcolors=blue]{hyperref}

\usepackage{caption}
\usepackage{subcaption}

\usepackage{booktabs}
\usepackage[table,xcdraw]{xcolor}

\title{Progress towards a megapixel linear-mode avalanche photodiode array for ultra-low background shortwave infrared astronomy}

\author[a]{Charles-Antoine Claveau}
\author[b]{Michael Bottom}
\author[a]{Shane Jacobson}
\author[b]{Klaus Hodapp}
\author[a]{Guillaume Huber}
\author[a]{Matthew Newland}
\author[a]{Aidan Walk}
\author[c]{Markus Loose}
\author[d]{Ian Baker}
\author[d]{Egle Zemaityte}
\author[d]{Matthew Hicks}
\author[d]{Keith Barnes}
\author[d]{Richard Powell}
\author[e]{Ryan Bradley}
\author[e]{Eric Moore}
\affil[a]{Institute for Astronomy, University of Hawai'i at M\=anoa, Hilo, HI 96720-2700, USA}
\affil[b]{University of Hawai'i at M\=anoa, Honolulu, HI 96822, USA}
\affil[c]{Markury Scientific Inc., Thousand Oaks, CA 91361, USA}
\affil[d]{Leonardo M.W. Ltd., Southampton, S015 0LG, UK}
\affil[e]{Hawaii Aerospace Corp., Honolulu, HI 96816, USA}

\authorinfo{Further author information:\\C.A.C.: E-mail: caclav@berkeley.edu}

\begin{document}
%\linenumbers
\maketitle

\begin{abstract}

Spectroscopy of Earth-like exoplanets and ultra-faint galaxies are priority science cases for the coming decades. Here, broadband source flux rates are measured in photons per square meter per hour, imposing extreme demands on detector performance, including dark currents lower than \mbox{1 e-/pixel/kilosecond}, read noise less than \mbox{1 e-/pixel/frame}, and large formats. There are currently no infrared detectors that meet these requirements. The University of Hawai'i and industrial partners are developing one promising technology, linear mode avalanche photodiodes (LmAPDs), which is on track to meet the above-mentioned requirements.

We present progress towards developing a science-grade, megapixel format linear-mode avalanche photodiode array for low background shortwave (1 - 2.4 um) infrared astronomy. Our latest results show outstanding performance, with dark current \textless 1e-4 electrons/pixel/second and read noise reducing by 30\% per volt of bias, reaching less than 1e-/pixel/frame in correlated double-sampling, and able to average down to $\sim$0.3 e-/pixel/frame when using multiple non-destructive reads. We present some on-sky data as well as comment on prospects for photon number resolving capability.
\end{abstract}


\section{Introduction} \label{sec:intro}
\footnote{portions of this introductory text adapted from \cite{claveau2022first}}

The high quantum efficiency (QE), low dark current (DC), and tunable cut-off wavelength of mercury cadmium telluride (HgCdTe) makes it the leading material for astronomical infrared detectors. Superb large format arrays such as the HAWAII family, manufactured by Teledyne imaging systems, are in regular use at observatories around the world, and comprise fifteen of the eighteen detectors on the James Webb Space Telescope. These arrays have dark current well below 1 e-/pixel/kilosecond when operated at temperatures below $\sim$60 K \cite{Regan2020}, and read noise of $\sim$10-15 e-/pix/frame, which may be reduced to $\sim$\mbox{5 e-/pix/frame} by frame averaging \cite{birkmann2018noise}. However, for photon starved science such as exoplanet imaging or faint galaxy spectroscopy, such arrays are too noisy\cite{Finger2004,Downing2008}. In particular, the read noise imposes a severe barrier, and has not been improved significantly in the three decades, as it is a fundamental limitation to the MOSFET-based source follower used in each readout pixel node \cite{kozlowski1998hgcdte}. There is no imminent path to overcoming this.

To put the noise in context, one may compare the relative contribution of dark current to read noise over reasonable frame times. For a total exposure time of $t$ and frame time $t_{fr}$, the variance of dark current is $DC \cdot t$ and that of read noise is $RN^2\cdot t/t_{fr}$. Assuming 5 e- of read noise and assuming 1 e-/pixel/kilosecond of dark current, the variance of read noise is 25 or 250 times higher for frame times of 1000 and 100 seconds. 

There is a pressing need for extremely low noise infrared detectors, as the latest astronomical decadal survey\cite{astrodecsurvey2020} identified a 6-meter space telescope operating from UV-IR the highest priority mission, with exoplanet imaging and spectroscopy of Earth-like exoplanets as the primary science driver. For this science, typical flux rates are about 1 photon per square meter per hour in V-band, and it is well known that detector noise is the most serious obstacle for such missions \cite{Robinson2020, Lacy2019, Crill2018}, leading to a need for dark currents \textless 0.001 e-/pix/s and \mbox{read noise \textless 0.3 rms e-/pix/frame}. In the optical, EMCCDs can meet these noise requirements, but in the infrared---where most of the deep biomarker spectral signatures exist---no current sensors are suitable. 

The benefits of such a sensor would also extend to ground-based astronomy. Modern high-resolution infrared spectrographs are designed with resolutions of 30,000 to 100,000, necessary to resolve stellar spectral or planetary lines. However, for spectral resolutions above $\sim$5000, all observations are read noise limited in the interline continuum \cite{sullivan2012calibrated}. As such, reducing the read noise is the most straightforward method of allowing such instruments to reach their full potential.

\subsection{Linear-mode avalanche photodiodes}
Linear-mode avalanche photodiodes (LmAPDs) offer one potential path to overcoming this read noise barrier. In these devices, large electric fields cause signal amplification through electron avalanching, so electrons are multiplied before the read noise penalty. This leads to an ``effective'' read noise, which is simply the base read noise divided by the multiplication gain\footnote{For example, operated at low bias voltage, before avalanching sets in, a received signal of 13 photoelectrons will encounter an underlying  read noise of $\sim$13 e-, giving a signal-to-noise ratio (SNR) of $\sim$1.  At moderate bias voltage, with an avalanche gain of $\sim$10, the signal of 13 electrons will be multiplied to $\sim$130 e-, so the SNR will be 10. This is equivalent to a 13 e- signal seeing an effective read noise of $\sim$13e-/10=1.3 e-.}. The price of this is a reduced full well, which is mostly irrelevant for the low flux rates in question.

HgCdTe LmAPDs, developed by Leonardo corporation (formerly Selex) in partnership with ESO and the University of Hawai'i, have found wide use as high speed wavefront sensors, such as in the SAPHIRA detectors.\cite{Finger2010} The current generation of SAPHIRA arrays have demonstrated sensitivity from 0.8 to 2.5 $\mu$m with high QE (\textgreater 80\%), fast pixel response, and an avalanche gain (aka APD gain) of \textgreater 500, offering an unmatched combination of sub-electron effective read noise (as low as 0.1 rms e-) at 1kHz frame rate at convenient operating temperatures of 90-100K\cite{Finger2016,Atkinson2018_2}. These detectors are now in regular use at major observatories around the world.

While SAPHIRA detectors can be used as science focal plane detectors, they are unsuitable for the low-flux science cases discussed above. First, their pixel format (320 x 256) is too small for integral field unit spectroscopy, which is better matched to a 4 megapixel (eg, 2k x 2k) sensor. Second, the lack of reference pixels puts high demands on voltage and temperature stability for long exposures, where drifts in the voltage level manifest as an extra noise source. Finally, when operated at low APD gain, SAPHIRAs deliver a dark current close to the level required, and still likely glow limited\cite{Atkinson2017}. However, when operated at high APD gain, to lower the effective read noise, the large voltages applied results in trap-assisted tunneling of electrons across the p-n junction, which causes the effective dark current to exponentially increase to unacceptable levels. In practice, this means that SAPHIRAs cannot deliver low dark current and low read noise simultaneously.

\begin{figure}[htbp]
  \centering
  \includegraphics[width=\columnwidth]{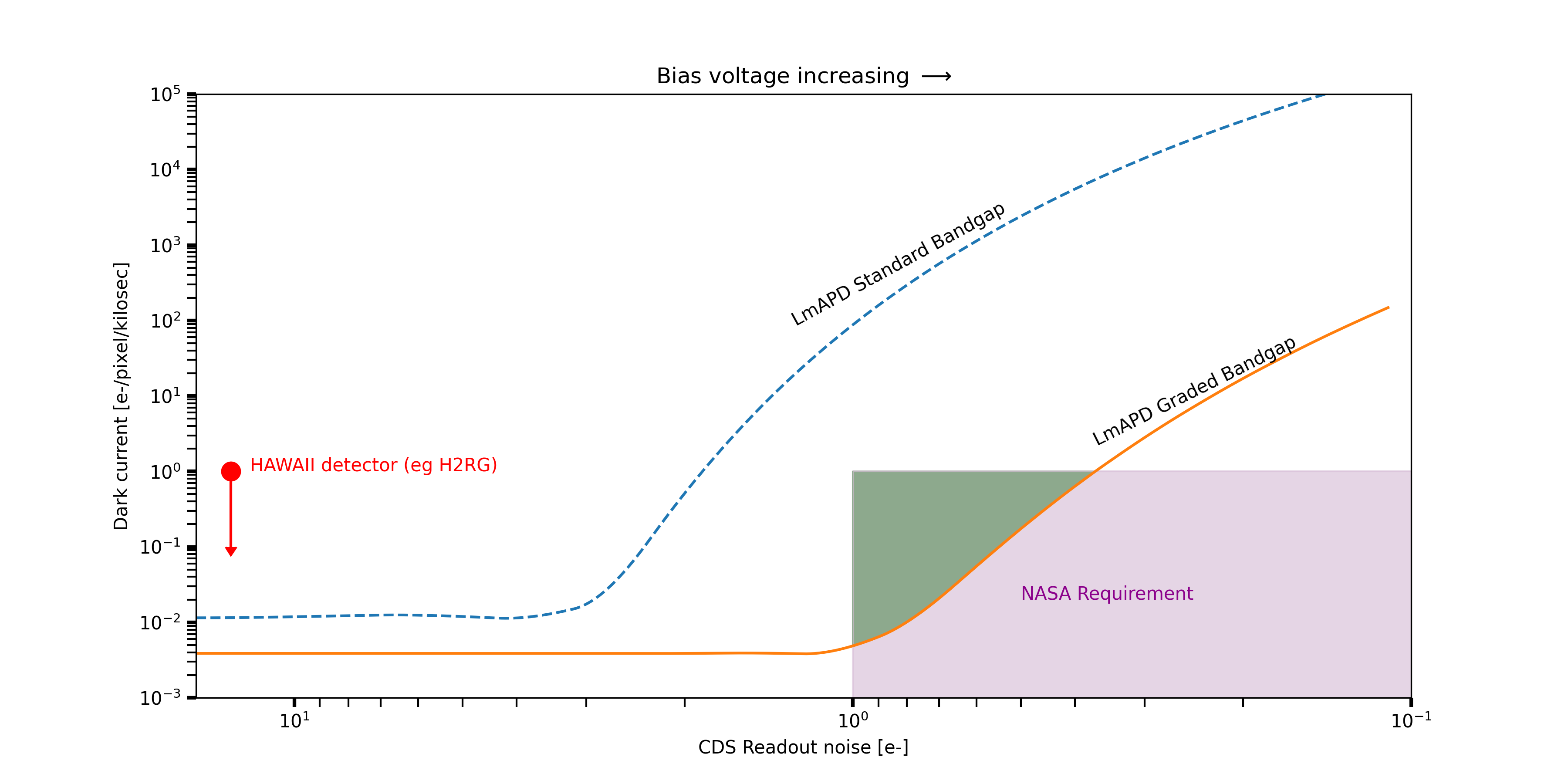}
  \caption{Theoretical performance curves of LmAPD bandgap designs. Graded-bandgap LmAPD designs can deliver the simultaneously low read noise and dark current needed for the next generation of NASA space missions (green shaded region). The bias voltage (upper x axis) controls the  read noise (lower x axis, decreasing to the right). The standard bandgap design has tunneling current at too low bias voltage (or too high read noise), after which the dark current increases exponentially. At low read noise, a graded bandgap design has dark current 1000-10,000x lower than a standard bandgap design.  These theoretical curves assume an operating temperature of 60K.}
  \label{fig:dc_vs_rn_vs_bv_v2}
\end{figure}

Fig.~\ref{fig:dc_vs_rn_vs_bv_v2} shows theoretical performance curves of the SAPHIRA bandgap as a dashed blue line. The effective dark current stays very low, and as the bias voltage increases (top axis, unlabeled), the  read noise reduces up until $\sim$3 e-. At that point, tunneling current begins to dominate the dark current budget, and by the time the read noise reaches 1 e-, it has increased by four orders of magnitude.

Motivated by the SAPHIRA results, the University of Hawai'i has partnered with Leonardo corporation, Markury Scientific, and Hawaii Aerospace to develop an LmAPD device suitable for ultra-low background infrared astronomy, with the goals of a dark current \textless 0.001 \mbox{e-/pix/s} and an effective read noise \textless 1.0 rms e-/pix/frame. The main differences from the SAPHIRA will be a larger format (1k x 1k with 15 $\mu$m pixels), a design including reference pixels to improve overall stability, and careful bandgap engineering to move the onset of tunneling current to higher voltage, so low read noise and dark current can be simultaneously achieved (see Fig.~\ref{fig:dc_vs_rn_vs_bv_v2}). 

Our development plan includes three stages. The first is fabricating and testing engineering-grade megapixel sensors with the same bandgaps as the SAPHIRAs, but with smaller pixels and reference pixels. These first devices are mainly aimed at proving the capabilities of making larger-format devices, but are not expected to push the state-of-the-art in sensitivity. Second, fabricating and testing larger science-grade sensors with a modified, graded bandgap, that should give low dark current and read noise simultaneously. Finally, radiation testing of the science-grade devices to relevant space-like levels.

The first stage, fabrication and testing of the engineering-grade sensors, took place in 2020-2021. Test results were quite encouraging, showing a dark current below 1e-/1000s at 50 K, a device glow of about one electron every 12 frames, and  read noise in correlated double-sampling decreasing by 30\% per volt following theoretical predictions, and reaching 2e- at 8 volts of bias. This work is reported in Ref.~\citenum{claveau2022first}. 

This paper presents the results obtained from the science grade versions of this new 1kx1k detector design. We describe the detector design, readout chain, and laboratory test environment. We present measured dark current, glow, and read noise, as well as a first attempt to demonstrate photon number resolving capability. We also discuss persistence and non-uniformity, and present on-sky data.

\section{Detector design and architecture}

\subsection{LmAPD technology in HgCdTe material}
\label{sec:detdesign:hgcdte}
The arrays are made up of HgCdTe layers grown by Metal Organic Vapour Phase Epitaxy (MOVPE)\cite{Maxey2010}. This manufacturing technique allows for a precise control of the semiconductor profile, so complex bandgaps and doping concentrations can be produced to achieve the desired structure and properties of the photodiode. HgCdTe is a material of choice among semiconductors for APD\cite{Baker2010,Beck2014} for the following reasons:
\begin{itemize}
    \item Only electrons participate in the avalanche process, because the holes have low mobility due to their high effective mass and low ionization efficiency.
    \item The avalanche process is ballistic because the electrons do not experience phonon interactions or scattering, resulting in nearly noiseless amplification and deterministic APD gain as a function of bias voltage.
    \item There is no breakdown effect in HgCdTe at high bias voltages/APD gains, as it is the case for Geiger mode photodiodes.
\end{itemize}

\begin{figure}[htbp]
  \centering
  \includegraphics[width=\linewidth]{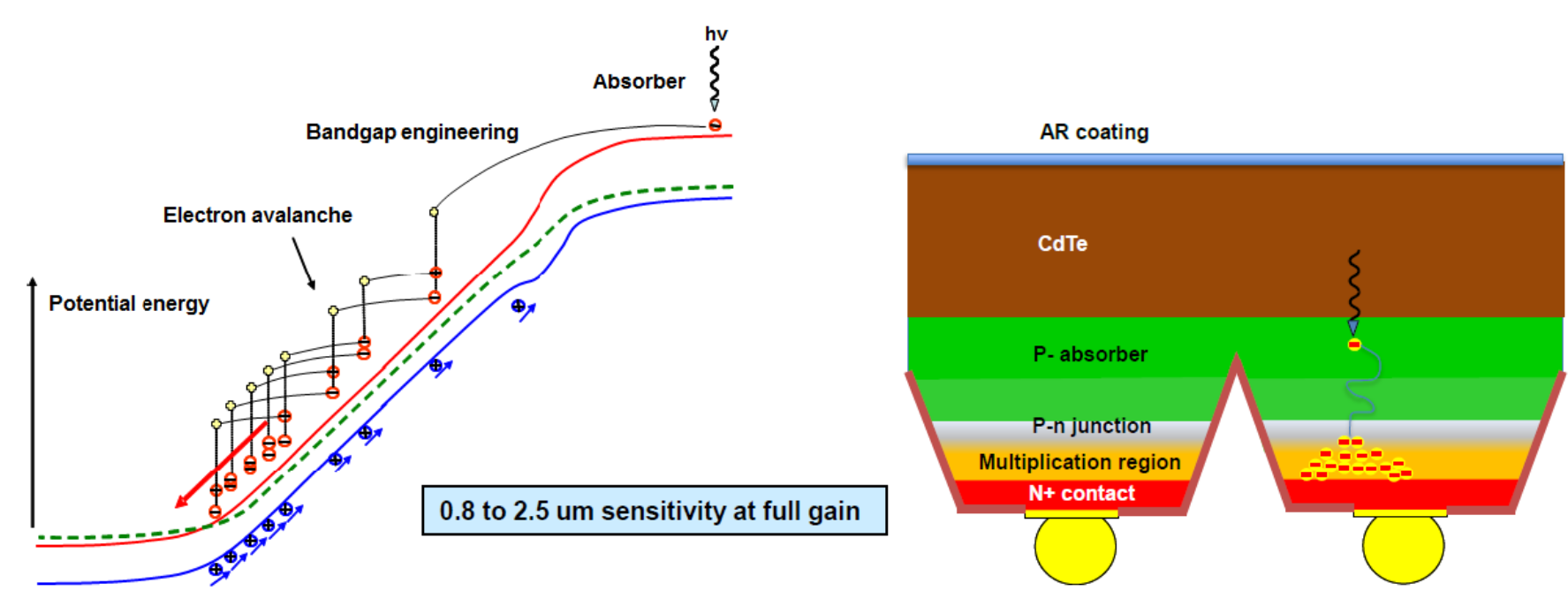}
  \vspace{\baselineskip}
  \caption{On the left, the single-carrier electron avalanche process is illustrated. The bias voltage is applied across the multiplication and junction regions. The electrons (red circles) experience the avalanche process in the multiplication region due to the applied electric field, while the positively charged holes (blue circles) remain stationary. On the right, a schematic of the APD heterostructure cross-section is shown. The ``yellow spheres'' refer to the indium bumps utilized to connect each pixel node with the underlying silicon ROIC - not shown here. Original figure courtesy of Leonardo corporation.}
  \label{fig:movpe_hgcdte_diode_design} 
\end{figure}

The top layer of the photodiode is a CdTe seed layer opaque at $\lambda $\textless0.8 $\mu$m, see Fig.~\ref{fig:movpe_hgcdte_diode_design}. Photons of \mbox{$\lambda \leqslant$2.5 $\mu$m} are absorbed in the p-type absorber directly grown on CdTe and are converted into electrons. The photon generated charge diffuses to the p-n junction and is then accelerated in the electric field of the multiplication region to start the avalanche process by impact ionization. The absorber and the multiplication region are therefore decoupled so each can be optimized separately. Photons with longer wavelengths ($\lambda$\textgreater 2.5 $\mu$m) can penetrate the absorption layer and the junction region to be absorbed directly in the multiplication region.

The array is hybridized to the silicon readout circuit (ROIC) via the contact pads and indium bumps embedded at the tip of each pixel mesa cone, see Fig.~\ref{fig:movpe_hgcdte_diode_design}. The mesa structure on the back of the device fulfill several critical functions. Their slots extending through the absorber layer prevent lateral charge collection to provide enhanced electrical isolation between pixels. Their shape maximizes photon trapping within the pixel to avoid stray photons from scattering into neighboring pixels. Together with a continuous absorber layer, this photodiode layout provides complete photon absorption within the pixel, and ensures every photoelectron receives the same APD gain independent of the position where the photon is absorbed in the pixel, so that no lenslets at the top of the pixels are needed. These three feature help to prevent any crosstalk effects between pixels that could degrade the modulation transfer function of the detector.

\subsection{Performance optimization and bandgap engineering}
\label{sec:detdesign:dark current}

Dark current (DC) refers to the generation of electrons within the semiconductor material itself. These electrons accumulate over the exposure time and thus add to the signal measured at the detector output, and have units of e-/pixel/second. In other words, the dark current can be considered as an excess signal that competes with or contaminates the target signal to be measured of photoelectrons coming from the integrated light flux. In addition, for a fixed integration time, the number of accumulated DC charges is a random variable that follows a Poisson distribution. Therefore, the DC is also a source of noise, similar to the read noise, although the latter is generated upstream during the readout of the pixel nodes by the ROIC - read noise is commonly modeled as a Gaussian random variable, centered around zero, with a variance equal to the read noise level.

In HgCdTe LmAPD arrays, the generative process of dark current which emanates primarily in the narrow bandgap multiplication region consists of two main mechanisms:

\begin{enumerate}
    \item \textbf{Thermal DC}, which increases with temperature and does not depend on the bias voltage applied to the photodiodes.
    \item \textbf{Trap-assisted tunneling current}, through which electrons cross the bandgap of the material in the multiplication region via trap states (defects or impurities) located within the bandgap. These traps serve as proxies for the charges to migrate from the valence band to the conduction band. Trap-assisted tunneling current occurs in two ways:
    \begin{itemize}
        \item In the case of \textbf{tunnel-tunnel current}, an electron tunnels directly through a trap state without requiring thermal excitation.
        \item In the case of \textbf{thermal-tunnel current}, an electron is first thermally excited into a trap state before tunneling through the bandgap.
    \end{itemize}
    At temperatures $\leq$ 65K, only tunnel-tunnel current is significant. The tunneling current is highly dependent on the level of impurities in the material, requiring very strict control over the quality of the semiconductor manufacturing process. Additionally, the tunneling current increases exponentially with high bias voltage, as a high electric field exerts a larger force on the electrons.
\end{enumerate}

As shown by the orange curve Fig.~\ref{fig:dc_vs_rn_vs_bv_v2}, the theoretical evolution of the DC as a function of the bias voltage, or APD gain, exhibits two regimes, here illustrated for a fixed temperature of T=50K. At low and moderate APD gains, the DC is dominated by the thermal component, which as expected remains constant with respect to the bias voltage. Then, upon crossing the voltage threshold, the onset of tunneling current occurs, becoming the dominant component of the DC. When the device is operated at high bias voltage, hence low effective read noise, the DC increases to unacceptable levels, rising exponentially.

Therefore, the key goal in designing the LmAPD arrays is to move up the voltage threshold so high avalanche gain can be achieved before tunneling current emerges and dominates the DC budget. This is accomplished by carefully engineering the bandgap. Firstly, the bandgap of the multiplication region has been widened to hinder the bandgap crossing. Secondly, a graded bandgap profile has been implemented in the multiplication region, which is expected to reduce the APD gain experienced by the thermally generated DC, and to allow for APD gains of several tens before the onset of trap-assisted tunneling current. As a result, this leads to another possible advantageous outcome. Thermal dark current is anticipated to experience a lower avalanche gain than photelectrons, which can be very useful for distinguishing between them, particularly under low-flux conditions. While the engineering-grade sensors already featured this widened bandgap, the science-grade devices discussed in this paper are the first to capitalize on this novel graded bandgap.

\subsection{Readout circuit characteristics}
\label{sec:detdesign:roicme1070}

\begin{table}[htbp]
\centering
\begin{tabular}{|l|ccc|}
\hline
 &
  SAPHIRA &
  \begin{tabular}[c]{@{}c@{}}ME1070\\ (engineering-grade, \\ already tested)\end{tabular} &
  \begin{tabular}[c]{@{}c@{}}ME1070\\ (science-grade,\\ this work)\end{tabular} \\ \hline
Pixel size                 & 24 $\mu$m & 15 $\mu$m & 15 $\mu$m         \\
Format                     & 320x256 $\text{pixs}^2$ & 1024x1024 $\text{pixs}^2$   & 1024x1024 $\text{pixs}^2$ \\
Max frame rate             & 1000 Hz & $\sim$10 Hz & $\sim$10 Hz        \\
Reference pixels           & No      & Yes         & Yes                \\
Number of video outputs    & 32      & 16          & 16                 \\
Bandgap structure          & Widened & Widened     & Widened and graded \\
Onset of tunneling current & 8 volts & 8 volts     & 13 volts (est.)     \\ \hline
\end{tabular}
\vspace{\baselineskip}
\caption{\label{tab:comp}Comparison of detector properties between SAPHIRA and 1kx1k LmAPD engineering and science-grade sensors.}
\end{table}

The ROIC is a low operating voltage (\textless3.5V) multiplexer custom-designed by Leonardo for the LmAPD array. Its current version name is ME1070. It has a small integration node capacitance per pixel of about 27 fF to limit read noise. All the I/O pad are grouped on one side of the chip in order to make the arrays 3-side buttable.

The device features 16 video outputs able to run at a pixel clock up to 1MHz/output channel, resulting in a maximum full-frame readout speed of about 15 frames per second (FPS). The 16 parallel video channels are organized in such a way that they read out 16 adjacent pixels simultaneously with a single ADC (Analog to Digital Converter) conversion strobe, so that windowing can be as small as 1 row by 16 columns. Scanning modes allowing for multiple non destructive readouts at the pixel level, or on a row by row basis are also possible. This facilitates multisampling beneficial to further minimize the read noise. The detector can be read out using the readout scheme called integrate then read (ITR) in conjunction with up-the-ramp (UTR) sampling; an external strobe to reset the full array is applied at the start of each new exposure. The readout scheme called read-reset-read for efficient correlated double sampling is also available.

The ROIC architecture has 4 reference rows of 1024 pixels each which can be read out either in interlaced mode, or as four extra rows added at the end of the read frames. This last mode is used in our tests. A reference pixel in NIR arrays is a pixel featuring the same capacitance as the rest of the sensitive pixels, but with the HgCdTe material removed, and is not exposed to light. They are useful to correct for low frequency noise that can arise from voltage and temperature drifts which can happen over long exposure times, e.g., when measuring very low dark current over hours or days. The ROIC architecture also has 4 reference outputs coupled to a row of 256 reference pixels to further compensate for any disturbance effects, such as electromagnetic interference (EMI).

\section{Laboratory testbed}
\subsection{Readout chain}
\label{sec:labsetup:readout}

\begin{figure}[htbp]
  \centering
  \includegraphics[scale=0.25]{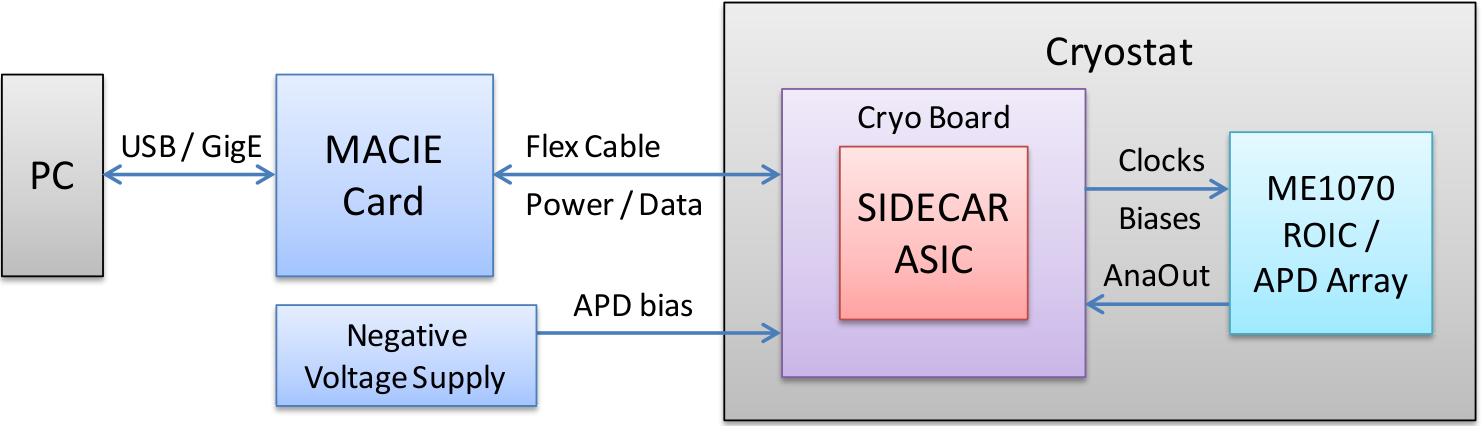}
  \vspace{\baselineskip}
  \caption{Block diagram of the control system and readout chain of the detector. See text for description.}
  \label{fig:rdout_chain_blkdiag}
\end{figure}

As shown on the diagram Fig.~\ref{fig:rdout_chain_blkdiag}, the main components used to control and read out the detector from a computer are:
\begin{enumerate}
    \item The detector is controlled by a SIDECAR ASIC (Application-Specific Integrated Circuit), located outside and just below the inner sanctum of the detector, and connected to it via an rigid flex cable supplied by Hawaii Aerospace. It is cooled down to the same temperature as the detector, being attached to the same cryogenic stage. The SIDECAR is a space-qualified microcontroller-based system-on-a-chip\cite{Loose2003} consisting of low-noise analog circuits, detector output conditioning including amplifiers, and 36 parallel ADC channels for 16-bit signal digitization. It can be operated both at ambient temperature and at cryogenic temperatures down to 40K.
    \item The SIDECAR is connected to a MACIE (Multi-purpose ASIC Control and Interface Electronics) controller card installed outside the cryostat using a flex cable routed through the wall of the test dewar. The MACIE card serves as the interface with the computer linked to it with a USB cable.
    \item The bias voltage that determines the APD gain of the detector is provided and adjusted by an external negative voltage referred to as the COMMON voltage. This bias voltage is given as the difference between the Pixel Reset Voltage (PRV) and the COMMON voltage, with PRV being set and delivered by the SIDECAR ASIC.

\end{enumerate}

We developed custom firmware and software to communicate with the detectors using the SIDECAR and MACIE cards. This includes a custom Cython-based Python wrapper for the MACIE C library, simplifying automation and testing. In principle, it is compatible with any detectors using SIDECAR ASIC or the next generation ACADIA ASIC\cite{Loose2018}, as they share the same unified MACIE interface.

\subsection{Cryostat and optical setup}
\label{sec:labsetup:cryostat}

\begin{figure}[htbp]
  \centering
  \includegraphics[width=\linewidth]{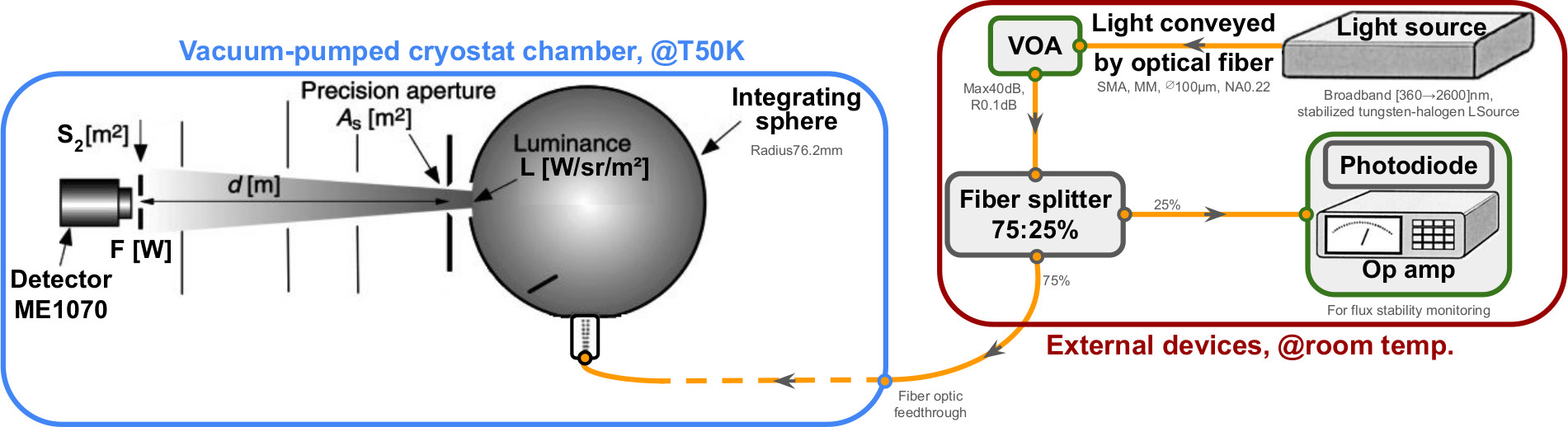}
  \vspace{\baselineskip}
  \caption{Diagram of the full test setup made up of the external illumination setup kept at room temperature (shown in the red box), combined with the cryogenic testbed (show in the blue box). The devices in the green bow are remotely controlled. See description in the text for more details.}
  \label{fig:test_setup_diagram}
\end{figure}

\begin{figure}[htbp]
  \begin{subfigure}{0.45\columnwidth}
    \includegraphics[width=\linewidth]{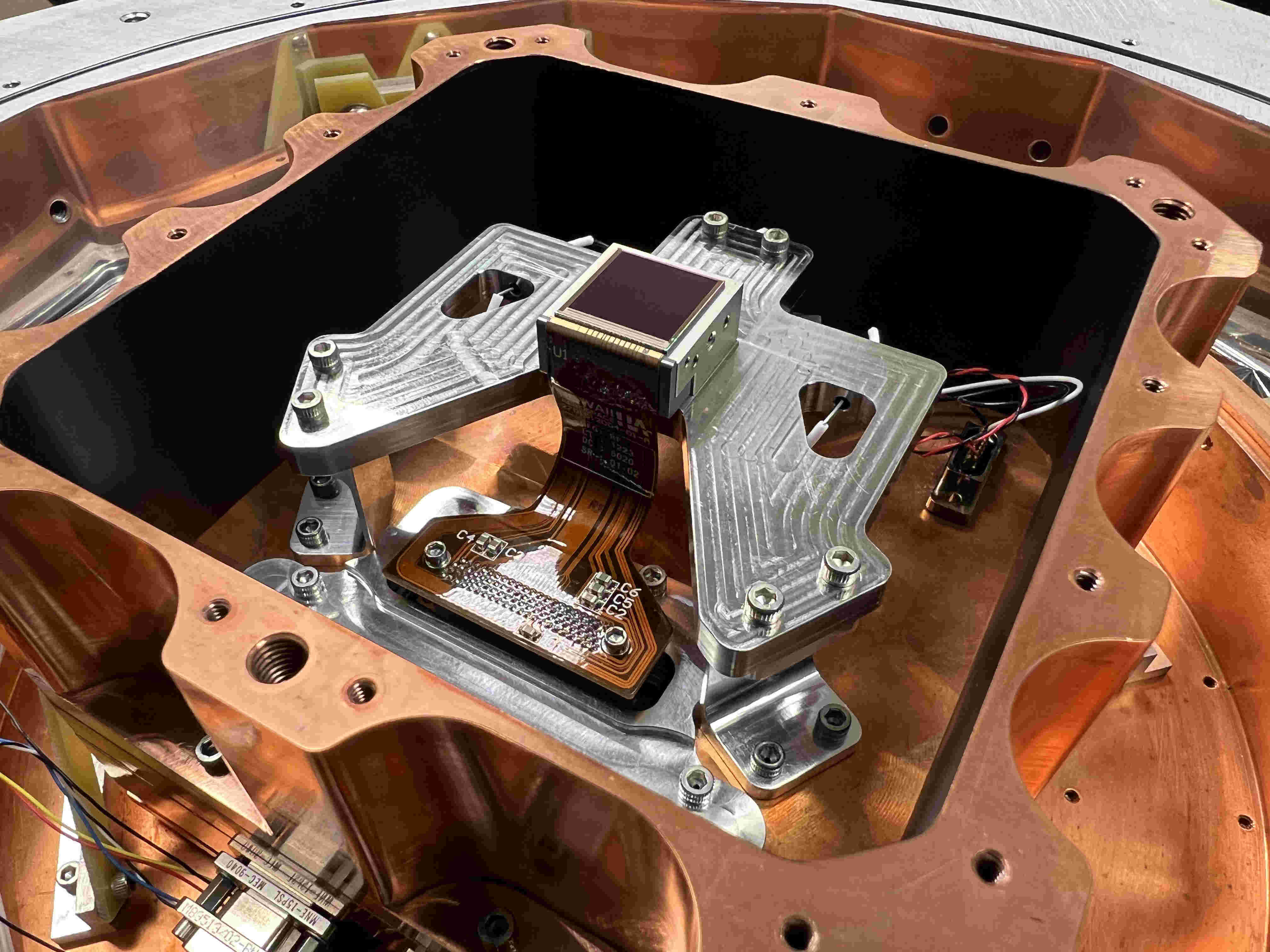}
    \caption{The inner detector housing and mounting structures. A 1kx1k LmAPD detector is integrated on its carrier and connected to the flex cable plugged to the SIDECAR ASIC located underneath the mounting site.}\label{fig:inner_det_housing}
  \end{subfigure}\hfill
  \begin{subfigure}{0.45\columnwidth}
    \includegraphics[width=\linewidth]{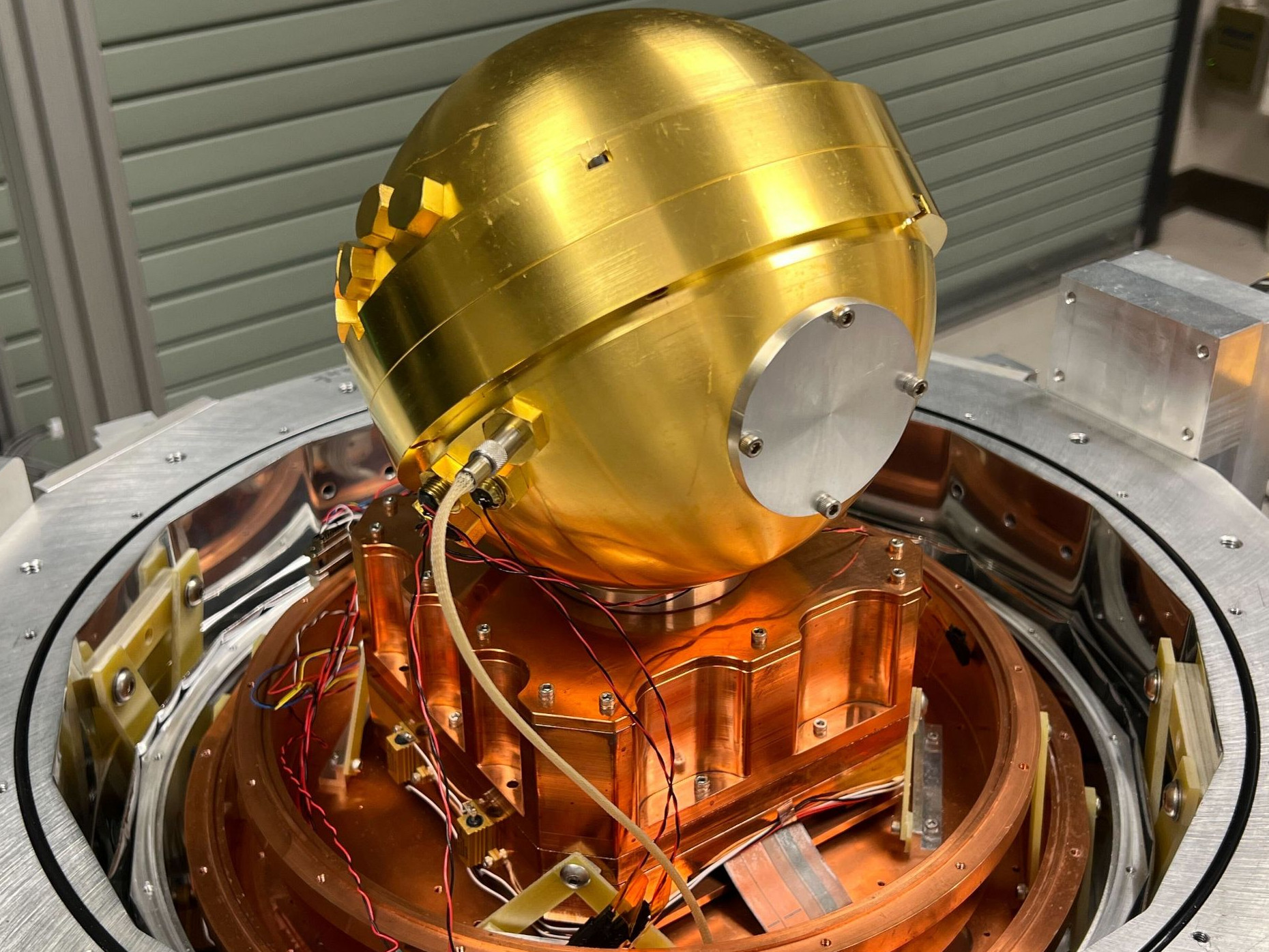}
    \caption{Gold integrating sphere mounted on top of the detector. The optical fiber is fitted in one port of the integrating sphere, visible just below the central seam.\newline}\label{fig:integrating_sphere}          
  \end{subfigure}\newline
  \caption{Pictures showing the cryogenic opened of ULBcam and the detector mounting site.}
  \label{fig:testbed}
\end{figure}

The test setup used to characterize and evaluate the performance of the science grade sensors is shown in the system diagram Fig.~\ref{fig:test_setup_diagram} and consists of two parts:

\begin{itemize}
    \item The illumination setup has been improved compared with the previous system used for the eng. grade sensors. It is designed to provide enhanced stability and control of the detector's illumination vs. time and temperature. To achieve this, the new illumination setup is entirely installed outside the cryostat, and kept at room temperature. In addition, its optical components are fully interconnected using optical fibers (multimode, 100 $\mu$m core diameter fiber) to convey the light. The light source (LS) is a stabilized tungsten-halogen LS (Thorlabs SLS201L) with an emission spectrum covering the spectral range from 360 to 2600 nm, close to a blackbody spectrum with a peak wavelength of 1 $\mu$m. A passband filter (Thorlabs FBH-1312) centered around $\lambda$=1.3um and with a narrow optical bandwidth (FWHM, @-3 dB) of 12 nm is inserted in the filter holder of the LS. The output of the LS is fiber-coupled to a multimode and broadband VOA (Variable Optical Attenuator) which has a theoretical maximum attenuation of 40dB, repeatability and precision of 0.1 dB, and operation wavelength from $\sim$400 to 2500 nm. However, the VOA we acquired didn't meet the expected performance, with tested capabilities falling well below the manufacturer's specifications. For instance, it exhibited a maximum attenuation of less than 30dB, a highly non-linear response despite the provided calibration file, and an opening response time of $\sim$5 seconds compared to $\sim$500 ms specified in the datasheet. The modulated output flux from the VOA is then split into 2 separate channels using a fiber splitter (Thorlabs TM105R2S1B). Approximately 75\% of the flux is routed by an optical fiber via a feedthrough in the cryostat wall to be injected into the cryogenic integrating sphere placed in front of the detector, see Fig.~\ref{fig:integrating_sphere}. The remaining 25\% of the light flux is fed into an external fiber-coupled photodetector (Thorlabs PDA10CS2), which is used to monitor the light flux. A test monitoring the flux received by the external photodetector at room temperature in parallel with that received by the detector in the cryogenic chamber showed a consistent signal evolution between the two devices, and a flux stability measured over several days of the order of 3\%. In addition, the maximum flux measured in the central region of the detector is $\sim$350 e-/pix/s .
    \item The Ultra-Low Background infrared Camera\cite{hall20044k} (ULBCam) instrument, which functions as both a lab testbed and on-sky camera at the University of Hawai'i 2.2m telescope. To characterize the detectors, we operate ULBCam in a testbed configuration, where a light-tight cover replaces the field lens to minimize thermal background. The cryostat chamber vacuum is pumped to \textless 1e-6 Torr and can be cooled down to a temperature as low as 40K, with milliKelvin stability over week-long periods. The internal focal plane assembly is shown Fig.~\ref{fig:inner_det_housing}, see description in the legend.
\end{itemize}

It was found that the optical system did not provide a perfectly light tight seal, with signal being detectable at the level of a few photoelectrons per pixel per hour. To mitigate this, we installed a mask $\sim$1mm above the detector. This mask is at the same temperature as the detector and has minimal thermal emission in wavelengths of interest. It has a rectangular opening in the center leaving about 1\% of the sensor surface exposed. This allows measurements requiring both partial illumination and dark conditions to be made with the same system.

\section{Detector performance}

Unless otherwise stated, all the results presented in this section were obtained with the same engineering-grade sensor, cooled down to a temperature of 50K, and set up in source follower mode, with its minimum operating voltages.

\subsection{Conversion gain and read noise}

The same method used for the engineering grade sensors is applied to measure conversion gain and read noise at different bias voltages for the science grade sensor. It is based on the photon transfer curve (PTC) technique\cite{Janesick2007}. The reader can refer to the paper Ref.~\citenum{claveau2022first} for more details on its implementation and required data collection and reduction. 

The conversion gain and read noise are obtained by fitting the function in Eq.~(\ref{eq:ptc}) to the PTC data. The read noise ($RN$, here in ADU) returned corresponds to the CDS (Correlated Double sampling) read noise. The conversion gain $G_{conv}$ in ADU/e- corresponds to the inverse of the slope of the linear part of the PTC. It depends on the charge gain $g$, the APD gain $G_{avl}$, and the excess noise factor ($ENF$). The term $FPN$ stands for the fixed pattern noise term. ENF can bias the conversion gain estimate, and is indistinguishable from $g$ and $G_{avl}$ solely by PTC measurement. It is assumed to be equal to 1 in the rest of this Section. See Sec.~\ref{sec:pnr} for a first attempt to estimate this quantity and verify the validity of this assumption. \textbf{The normalized dark current and the effective read noise are directly derived by multiplying their raw value measured in ADU by the conversion gain, in order to convert them into e- and correct them for the APD gain.} Our PTC measurements at different bias voltages are reported Tab.~\ref{table:glob_perfs} and discussed Sec.~\ref{sec:perf:overall}.

\begin{equation}
\label{eq:ptc}
\sigma_{[ADU]} = \sqrt{RN^2 + \tfrac{1}{G_{conv}} \cdot N_{[ADU]} + \left( FPN \cdot N_{[ADU]} \right)^{2}},\qquad \text{with}\ G_{conv} = \frac{g}{G_{avl}\cdot ENF}
\end{equation}

\subsection{Interpixel capacitance}

CMOS sensors are affected by the interpixel capacitance (IPC)\cite{Finger2005}. IPC comes from the coupling between neighboring pixels' capacitance when pixels accumulate charge. This results in a reduction of the variance computed by simple spatial averaging. As a consequence, the linear slope of the PTC is reduced, which leads to an overestimation of the conversion gain and affects estimates of other quantities that depend on it, like the effective read noise and normalized dark current. Even sub-percent coupling coefficients can induce an error of the order of percent on the conversion gain.

\begin{figure}[htbp]
  \centering
  \includegraphics[width=0.6\linewidth]{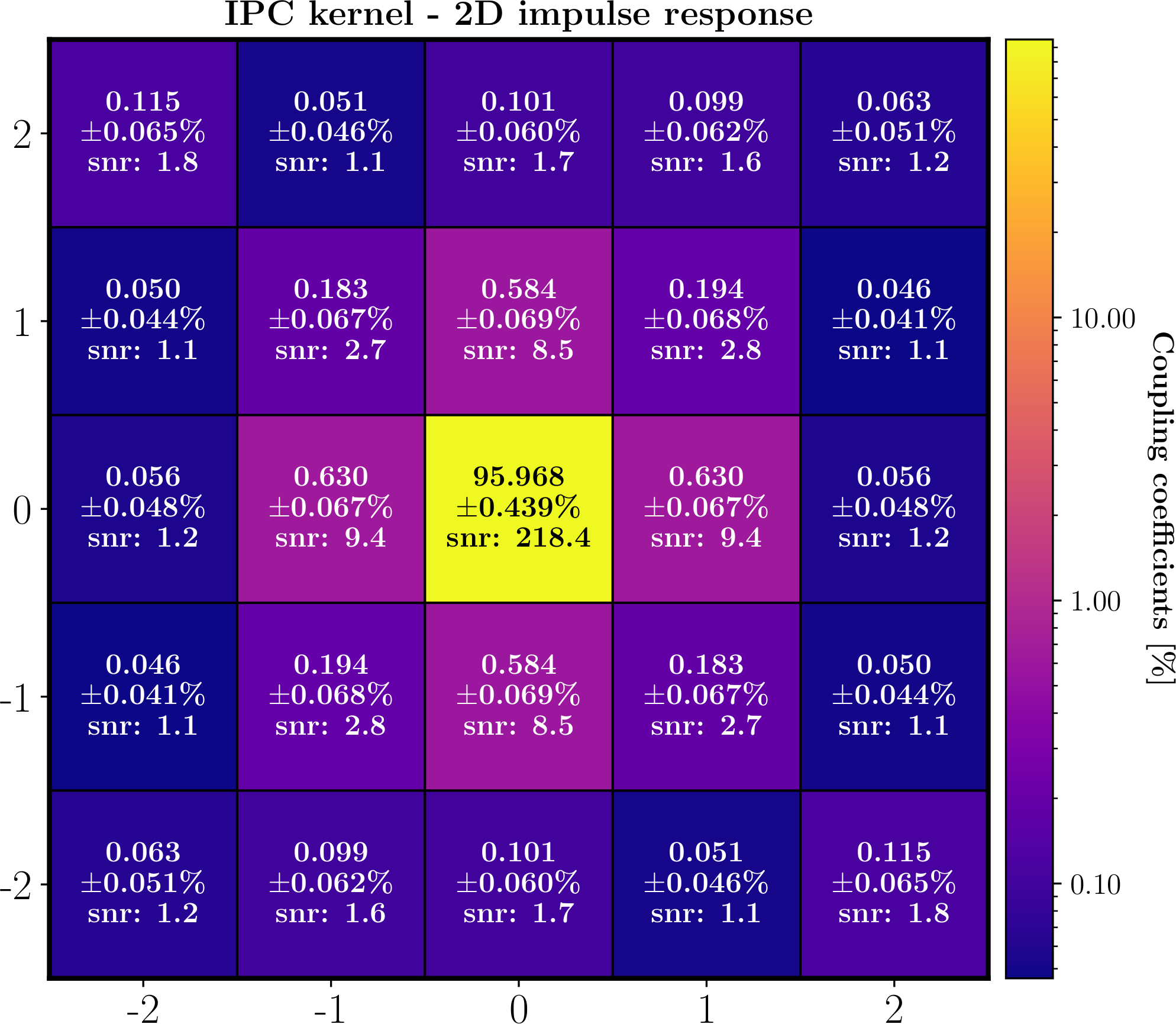}
  \vspace{\baselineskip}
  \caption{Measured 2D IPC kernel at bias voltage of 5V. Its estimation is limited to 5x5 terms. The IPC kernel is assumed symmetric, so only 15 terms (of the upper or lower triangular matrix) are computed. For each kernel term, the top value is the mean value of its distributions sampled by MCMC, and the second value is the absolute STD of its distribution, taken as the associated uncertainty. Both values are given as percentages. The corresponding SNRs are given as the ratio of these two values.}
  \label{fig:ipc_krnl}
\end{figure}

The IPC was characterized using 2 methods:
\begin{enumerate}
    \item  The method described in Moore et al.\cite{Moore2006}, which recovers the correct variance using the 2D autocorrelation function applied to data obtained from the pixel-level subtraction of 2 flat field frames. The Moore method makes the approximation that most of IPC is in the 4 immediate neighbor pixels. A stack of 500 pairs of flat field frames of 100x100 pix$^{2}$ was taken at constant illumination to compute the required 2D autocorrelation matrix. Coupling coefficient, or “leak" of signal in each of the 4 adjacent pixels, is estimated at $\sim$0.7\%.
    \item A new method that we developed, since the Moore method has been shown to have some shortcomings in overcorrecting and adding noise~\cite{LeGra2022}, and doesn't return in its initial form uncertainties on the output parameters. The same data of subtracted flat fields is used. This is a Bayesian method considering each pair of neighboring pixels as a pair of correlated random variables following a Gaussian bivariate distribution centered around zeros and defined by its symmetric covariance matrix. The 2D autocorrelation function is replaced by fitting the covariance matrix on the 2 subsets of pixels associated with the autocorrelation term computed. In other words, the two equal off-diagonal terms of the fitted covariance matrix gives a direct measure of the corresponding off-central term of the 2D autocorrelation matrix. The central term is computed as the variance for a Gaussian likelihood fitted on the data. MCMC is used to estimate the uncertainties. The method does not approximate the form of the IPC kernel and the full IPC kernel is recovered by FFTs from the autocorrelation matrix, as described in Moore et al, with error propagation of the uncertainties.
    
    In fact, the autocorrelation terms are estimated at different increasing signal levels and the slopes of the terms vs. signal level are fitted with a linear regression and used to reconstruct the autocorrelation matrix. Indeed, as described in Moore et al., the method for determining the IPC kernel does not depend on the light signal received by the detector. In addition, all autocorrelation terms scale linearly with the received signal by the same factor. This difference from the initial method can be beneficial when the detector operates at low flux, for example to remain within its linear regime. Possible residual correlations, such as those due to electronic noise, that do not depend on the flux signal are captured by the intercept of the fitted line. These correlations could prevent the peripheral terms of the IPC kernel from approaching zero as expected when moving away from the central term. Note that the auxiliary kernel composed of the “intercept" terms could also be used to study these correlations.
    
    Four stacks of 50 pairs of flat field frames of 100x100 pix$^{2}$ were taken, the first one in dark conditions to constrain the noise, and the others at three increasing levels of integrated signal. The resulting 5x5 kernel is shown Fig.~\ref{fig:ipc_krnl}. The most significant terms of the IPC kernel are found in the 4 adjacent pixels with an average coupling coefficient of $\sim$0.6$\pm$0.07\%. The other kernel terms are consistent with zero given the uncertainties. The total sum of the kernel terms is 1.003 very close to the value of 1 expected by charge conservation.
\end{enumerate}
The two methods returns a compatible coupling coefficient given the uncertainties which gives a good confidence in its estimated value. The IPC result means that there is a systematic over-estimation of the conversion gain measured using the baseline PTC method of $\sim$4\%, as can also be read directly in the central term of the IPC kernel $\sim$96\%. This effect is not corrected, as it is minor.

\subsection{Dark current and glow}

As shown in the previous characterization of the engineering grade sensors\cite{claveau2022first}, the dark current signal measured in these devices consists of two components: 1- a “continuous/per-time" component which is the intrinsic dark current originating from the HgCdTe material, it is given in e-/pixel/second, and 2- a “per-frame” component which is due to the glow of the ROIC when the acquisition of a read frame is triggered, it is given in e-/pixel/read.

The method and data used for measuring dark current and glow are similar to those employed in Ref.~\citenum{claveau2022first}. The procedure outlined in Ref.~\citenum{Regan2020} is applied to separately determine glow and dark current based on frame diversity. For each tested bias voltage, two identical 3-hour integrations are taken in dark conditions and regularly sampled, the first one 489 times, and the second one 81 times.

The mean glow-subtracted dark current value reported is given by the peak of the pixel distribution and the uncertainty $\sigma$ on this value is given by the median absolute deviation of the distribution around the dark current peak. The pixels outside $\text{dark current}_{\text{peak}} \pm 3 \sigma$ are rejected as outliers. Dark current measurements at different bias voltages are reported Tab.~\ref{table:glob_perfs} and discussed Sec.~\ref{sec:perf:overall}. The mean estimated glow is 0.03 e-/pixel/read, which is half of what was previously measured with the engineering grade detectors. The increasing number of outlier pixels with anomalously high dark current as the detector bias voltage increases are due to the same mechanism as the base tunneling current, which occurs at lower bias voltage for those pixels.

The source of the glow photons has been identified to come from within the pixel nodes from the source follower MOSFETs, and trace back to the removal of a metallization layer meant to shield the glow. A new glow-suppressed chip with this metallization restored was manufactured and delivered for testing. Its glow properties are the subject of the paper Ref.~\citenum{Huber2024}.

\begin{table}[htbp]
\caption{Summary table for different bias voltages of the conversion gain, raw read noise, and effective read noise determined by PTC method, as well as glow-subtracted dark current and glow measured using the 3-hour baseline method. The column “Reduction factor" gives the ratio between the conversion gain of the previous bias voltage and the current one. The number of pixels rejected as outliers in the dark current tests is also reported, as well as the effective full well capacity (Eff. FWC). The Eff. FWC is equal to the theoretical maximum dynamic range per pixel of 16 bits, i.e. $2^{16}$ levels in ADU, multiplied by the conversion gain. The row for bias voltage=15 volts is shaded in gray because PTC measurement was not possible data was too noisy for this bias voltage. The DC field for a 14V bias is also shaded because the glow and dark current could not be separated, as the tunneling current dominated the signal of the two datasets used for frame diversity.\newline}
\label{table:glob_perfs}
\resizebox{\textwidth}{!}{%
\begin{tabular}{ccccccccc}
\hline
\textbf{\begin{tabular}[c]{@{}c@{}}Bias\\ voltages {[}V{]}\end{tabular}} & \textbf{\begin{tabular}[c]{@{}c@{}}$\mathbf{K_{gain}}$\\ {[}$\mathbf{\textbf{e}^{-}/\textbf{ADU}}${]}\end{tabular}} & \textbf{\begin{tabular}[c]{@{}c@{}}Reduction\\ factor\end{tabular}} & \textbf{\begin{tabular}[c]{@{}c@{}}Raw RN\\ {[}ADU{]}\end{tabular}} & \textbf{\begin{tabular}[c]{@{}c@{}}Eff. RN\\ {[}$\mathbf{\textbf{e}^{-}}${]}\end{tabular}} & \textbf{\begin{tabular}[c]{@{}c@{}}DC\\ {[}$\mathbf{\textbf{e}^{-}/\textbf{pix}/ks}${]}\end{tabular}} & \textbf{\begin{tabular}[c]{@{}c@{}}Glow\\ {[}$\mathbf{\textbf{e}^{-}/\textbf{pix}/read}${]}\end{tabular}} & \textbf{\begin{tabular}[c]{@{}c@{}}\#Rejected\\ pixels {[}\%{]}\end{tabular}} & \textbf{\begin{tabular}[c]{@{}c@{}}Eff. FWC\\ {[}$\mathbf{\textbf{ke}^{-}}${]}\end{tabular}} \\ \hline
\textbf{3.0}                                                             & $1.32 \pm 0.05$                                                                                                     & N.A.                                                                & $7.78 \pm 0.40$                                                     & $10.3 \pm 0.8$                                                                             & $1.6 \pm 0.01$                                                                                        & $0.037 \pm 0.004$                                                                                         & 3.0                                                                           & 86.5                                                                                         \\
\textbf{4.0}                                                             & $1.16 \pm 0.02$                                                                                                     & 1.14                                                                & $7.55 \pm 0.20$                                                     & $8.8 \pm 0.3$                                                                              & $0.37 \pm 0.22$                                                                                       & $0.034 \pm 0.004$                                                                                         & 2.2                                                                           & 76.0                                                                                         \\
\textbf{5.0}                                                             & $0.95 \pm 0.03$                                                                                                     & 1.22                                                                & $7.34 \pm 0.30$                                                     & $7.0 \pm 0.4$                                                                              & $0.088 \pm 0.22$                                                                                      & $0.029 \pm 0.004$                                                                                         & 2.5                                                                           & 62.3                                                                                         \\
\textbf{6.0}                                                             & $0.71 \pm 0.01$                                                                                                     & 1.34                                                                & $7.58 \pm 0.20$                                                     & $5.4 \pm 0.2$                                                                              & $0.075 \pm 0.22$                                                                                      & $0.026 \pm 0.004$                                                                                         & 2.8                                                                           & 46.5                                                                                         \\
\textbf{7.0}                                                             & $0.55 \pm 0.01$                                                                                                     & 1.29                                                                & $7.8 \pm 0.30$                                                      & $4.3 \pm 0.2$                                                                              & $0.10 \pm 0.22$                                                                                       & $0.024 \pm 0.004$                                                                                         & 3.2                                                                           & 36.0                                                                                         \\
\textbf{8.0}                                                             & $0.40 \pm 0.005$                                                                                                    & 1.38                                                                & $7.47 \pm 0.25$                                                     & $3.0 \pm 0.1$                                                                              & $-0.20 \pm 0.22$                                                                                      & $0.032 \pm 0.004$                                                                                         & 2.7                                                                           & 26.2                                                                                         \\
\textbf{9.0}                                                             & $0.29 \pm 0.004$                                                                                                    & 1.38                                                                & $7.8 \pm 0.30$                                                      & $2.3 \pm 0.1$                                                                              & $-0.15 \pm 0.22$                                                                                      & $0.032 \pm 0.004$                                                                                         & 4.6                                                                           & 19.0                                                                                         \\
\textbf{10.0}                                                            & $0.21 \pm 0.02$                                                                                                     & 1.38                                                                & $7.9 \pm 0.30$                                                      & $1.7 \pm 0.2$                                                                              & $-0.16 \pm 0.22$                                                                                      & $0.025 \pm 0.004$                                                                                         & 6.7                                                                           & 13.8                                                                                         \\
\textbf{11.0}                                                            & $0.15 \pm 0.02$                                                                                                     & 1.40                                                                & $7.8 \pm 0.35$                                                      & $1.2 \pm 0.2$                                                                              & $0.18 \pm 0.22$                                                                                       & $0.028 \pm 0.004$                                                                                         & 16.3                                                                          & 9.8                                                                                          \\
\textbf{12.0}                                                            & $0.11 \pm 0.001$                                                                                                    & 1.36                                                                & $7.3 \pm 0.50$                                                      & $0.8 \pm 0.1$                                                                              & $0.39 \pm 0.22$                                                                                       & $0.026 \pm 0.004$                                                                                         & 31.2                                                                          & 7.2                                                                                          \\
\textbf{13.0}                                                            & $0.085 \pm 0.002$                                                                                                   & 1.29                                                                & $7.6 \pm 0.80$                                                      & $0.65 \pm 0.1$                                                                             & $2.93 \pm 0.03$                                                                                       & $0.029 \pm 0.004$                                                                                         & 30.6                                                                          & 5.6                                                                                          \\
\textbf{14.0}                                                            & $0.067 \pm 0.002$                                                                                                   & 1.27                                                                & $6.3 \pm 1.00$                                                      & $0.42 \pm 0.1$                                                                             & \cellcolor[HTML]{C0C0C0}/                                                                             & \cellcolor[HTML]{C0C0C0}/                                                                                 & 26.2                                                                          & 4.4                                                                                          \\
\rowcolor[HTML]{C0C0C0} 
\textbf{15.0}                                                            & /                                                                                                                   & /                                                                   & /                                                                   & /                                                                                          & /                                                                                                     & \cellcolor[HTML]{C0C0C0}/                                                                                 & 22.1                                                                          & /                                                                                            \\ \hline
\end{tabular}}
\end{table}

\subsection{Overall performance as a function of bias voltage}
\label{sec:perf:overall}

\begin{figure}[htbp]
  \centering
  \includegraphics[scale=0.70]{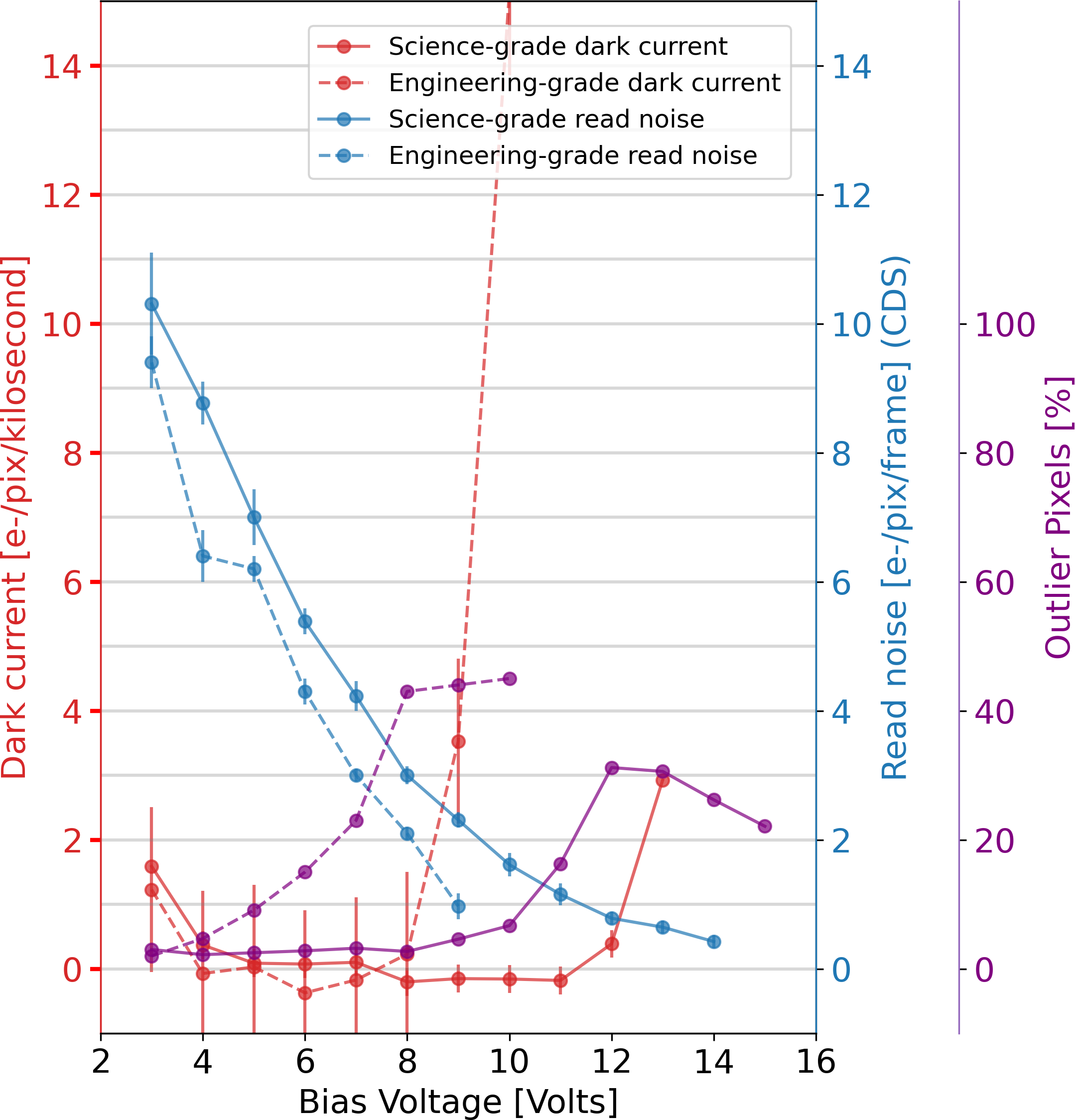}
  \vspace{\baselineskip}
  \caption{Measured normalized dark current and effective read noise as a function of bias voltage for the 1kx1k science-grade sensor, compared to best results previously obtained with the 1kx1k engineering-grade sensor (data collected from Ref.~\citenum{claveau2022first}). The proportion of pixels rejected as outliers in the dark current tests is also reported. A steep increase of the dark current starts from bias voltage \textgreater11V due to the onset of the tunneling current.}
  \label{fig:overall_perfs}
\end{figure}

The results of the joint analysis of dark current and readout noise as a function of bias voltage for the science and engineering-grade devices is shown Fig.~\ref{fig:overall_perfs}. The left axis shows the dark current, and the right axis show the readout noise in electrons, both on a linear scale. An additional right axis indicates the number of pixels rejected as outliers from the dark current distributions.

Fig.~\ref{fig:overall_perfs} shows several important effects:
\begin{enumerate}
    \item The readout noise behaves as predicted by theory, reducing by 30\% per volt of bias, consistent in both devices.
    \item The (glow-subtracted) dark current results are also the same between the two designs, at ~0.1 e-/pixel/kilosecond at low bias voltages. This was also verified using very long integrations (\textgreater 60 hours) at 5V and 10V of bias. At approximately 8 volts, the tunneling current begins to dominate the dark current budget for the engineering-grade device, and exponentially increases with voltage. This is pushed out to $\sim$11 volts for the science-grade device. 
    \item The percentage of outlier pixels exhibiting tunneling current (as a percent of total pixels, shown in purple) is greatly reduced in the science-grade devices. The engineering-grade sensor immediately begins to show increasing number of outlier pixels and shows an unacceptably high percentage by 6 volts. The science-grade detector has a small percentage of outlier pixels which do not increase to unacceptable levels until $\sim$11 volts.
\end{enumerate}

% and with SAPHIRA (data collected from Ref.~\citenum{Atkinson2017,Pastrana2018})

\subsection{Reduced readout noise by Fowler-sampling}
\label{sec:perf:red_ron}

The previous section reported readout noise in correlated double sampling (CDS), which is the simplest and most conservative way to report results. However, infrared detectors are almost always operated with more complex sampling schemes involving multiple nondestructive reads, as this causes the read noise to reduce by factors of 3-5\cite{birkmann2018noise}. The degree to which read noise averages down is important to determining the ultimate detector performance, and presents the fairest comparison between these devices and the current state-of-the-art.

\begin{figure}[htbp]
  \centering
  \includegraphics[width=0.7\linewidth]{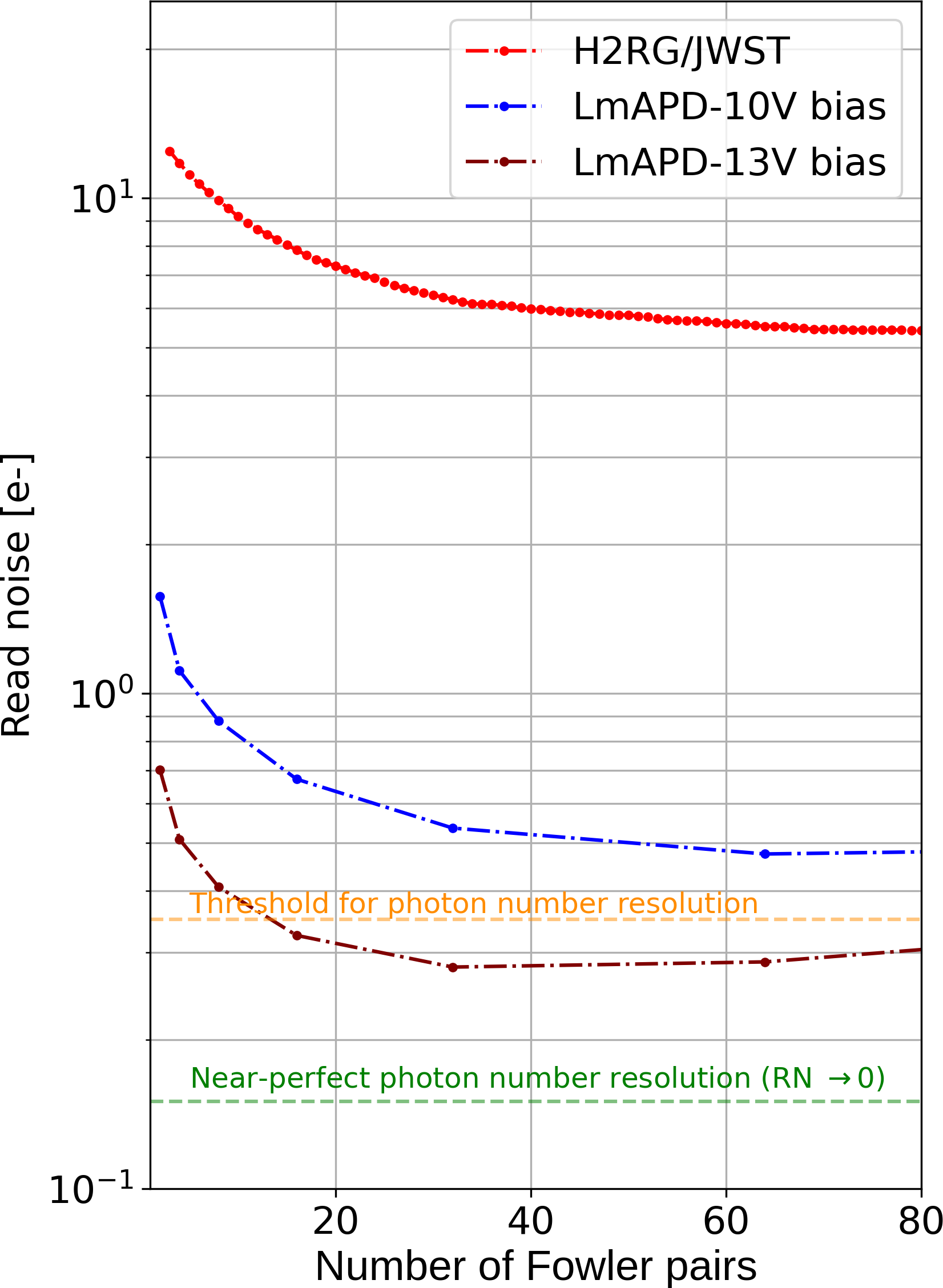}
  \vspace{\baselineskip}
  \caption{Reduction of the effective readout noise as a function of the number of frames averaged, based on the N-Fowler non-destructive multi-sampling technique. The results obtained for the 1kx1k LmAPD detector at two different high bias voltages, namely 10 and 13V, are compared to the values reported in Ref.~\citenum{birkmann2018noise} for the H2RGs on JWST. The green line shows the level of read noise ($\sim$0.15 e-) below which it is in principle possible to nearly perfectly resolve photon events.}
  \label{fig:rn_fwlr_sampling}
\end{figure}

Fig.~\ref{fig:rn_fwlr_sampling} shows the properties of read noise averaging following Fowler method. An identical dataset composed of 256 read frames acquired by UTR-sampling at maximum readout speed (1.45FPS) is repeated 25 times. The reduced read noise is computed at the pixel level by temporal averaging using the 25 datasets and the median value of the pixel distribution is reported for different numbers of Fowler frame pairs. For instance, in this test, a 30-Fowler sampling means that an averaged stack of 30 consecutive read frames taken at the beginning of an UTR-sampled exposure is subtracted from an averaged stack of 30 other read frames taken within the same exposure, the 2 frame subsets being also consecutive. In order to avoid the test being limited by the per-frame glow, it was performed in a mode where the pixel node source follower current is lowered by a factor x4, and PRV (pixel reset voltage) is slightly increased by +100 mV (decreasing the delta voltage between the gate and source terminals of the pixel source follower MOSFET), resulting in a total reduction of the glow by a factor of 14. The read noise reaches a minimum with about $\sim$30 pairs of frames averaged. At 10V bias, arguably the best bias for the device (see Fig.~\ref{fig:overall_perfs}), the read noise reaches a minimum of $\sim$0.5 e-, and at 13V a minimum of \textless0.3e-. In the latter case, such a low read noise should make possible for individual photon events to be distinguished, though an additional factor of $\sim$2 is needed before they can be perfectly distinguished and read noise is effectively zero. Both cases represent at least an order of magnitude improvement over the state-of-the-art H2RGs. The reduction of the read noise vs. number of averaged frames eventually flattens. This is commonly attributed to residual correlated read noise that does not average out. In addition, the curve for a 13V bias rises again for a high number of frames averaged, probably because of the glow which, though reduced, is still present and amplified at high bias, accumulating alongside the tunneling current.

\section{Photon number resolving and excess noise factor}
\label{sec:pnr}

\subsection{A few words about the context}

Building on previous results showing that the effective readout noise can reach a level as low as $\sim$0.30 e- at a bias voltage of 13V by combining a fairly high multiplication gain with multisampling and averaging, an initial attempt was made to demonstrate whether the detector is capable of photon number resolving (PNR). To our knowledge, this has never been demonstrated before for a megapixel detector in the NIR. Photon counting typically refers to the limited capability to distinguish only two scenarios: either 0 photons are received, or at least one photon - or more - is detected, EMCCDs are photon counting devices. Conversely, in the case of PNR, it is possible to count multiple photons in a single frame, which theoretically makes it possible to perform temporal or spatial noiseless pixel rebinning a posteriori, and to get very high dynamic range\cite{Gach2022}.

The commonly used statistical tool is the PNR histogram\cite{Hendrickson_2023_1, Hendrickson_2023_2}. It is constructed from several thousand, or more, PNR samples corresponding to the measured signal for a single exposure, generally at low flux, from a single pixel. Visually, it is possible to recognize and distinguish the individual peaks in the PNR histogram associated with the discrete arrival of photon packets.%, as governed by Poisson's Law.
%The idea is not new, as evidenced by the growing number of papers on this topic in recent years, particularly motivated by recent advancements in visible CMOS sensors\cite{Ma2022}. These detectors are equipped with very small and low capacitance pixels, and offer a read noise below $\sim$0.25 e-. Some of those papers present new methods to maximize the extraction of information from the PNR histogram for accurate calibration of every pixels\cite{Hendrickson2024}. Others introduce new operating strategies, such as reading out the detector as quickly as possible in a low-bit format, e.g. binary images, and then how to best reconstruct the final image from the generated data\cite{Chan2023}.

\subsection{Statistical model and analysis method}

Analysis of the PNR histograms is conducted by modeling the generative process of the signal measured in ADU at the output of a pixel, i.e., by determining the likelihood of the data. In the simplest case, the signal is first generated by a (discrete) Poisson distribution defined by its mean equal to the mean integrated flux in photoelectrons received by a pixel. It is then multiplied by the inverse of the conversion gain in e-/ADU. Finally is added the read noise commonly modeled by a Gaussian distribution with zero mean and a sigma equal to the read noise in ADU. This is written as a Poisson distribution, scaled by the inverse of the conversion gain, then convolved with a Gaussian distribution. In other words, the resulting distribution is a mixture distribution made up of an infinite number of identical Gaussian distribution components which are weighted by the Poisson distribution, and shifted by an integer multiple of the inverse of the conversion gain. The first peak is centered at 0 ADU. Each consecutive peak is spaced by the same distance equal to the inverse of the conversion gain.

A second objective of the PNR test is to determine the excess noise factor (ENF), if any, of the detector. The ENF arises when the multiplication gain is not perfectly deterministic but stochastic, introducing an additional source of noise in the output signal. In this case, the multiplication gain is no longer defined by a fixed value but by a probability distribution. In this study, a Gamma distribution is adopted, as motivated by the study in Ref.~\citenum{Lanthermann2019} where the ENF of a SAPHIRA detector is investigated. The ENF is defined by the formula Eq.~(\ref{eq:enf}) below, where the parameters $\left(\mu_{avlch.gn}\right)$ and $\left(\sigma_{avlch.gn}\right)^{2}$ are respectively the mean and variance of the APD gain distribution. For a perfectly deterministic APD gain, the ENF equals 1:

\begin{equation}
\label{eq:enf}
ENF = 1 + \frac{\left(\sigma_{avlch.gn}\right)^{2}}{\left(\mu_{avlch.gn}\right)^{2}}
\end{equation}

An important point is that each photoelectron converted in the photodiode actually experiences the avalanche process independently, so a different APD gain in the case of a non-deterministic amplification. Consequently, for a given component numbered $n$ of the mixture distribution, corresponding to the peak of the Poisson distribution associated with $n$ received photoelectrons, this component is the convolution of the Gaussian distribution with the Gamma distribution convolved with itself $n$ times. Lastly, the Gaussian distribution of the read noise can also have a non-zero mean, providing additional flexibility to the model to accommodate potential slight shifts or offsets of the overall distribution. In summary, the model consists of 5 PNR parameters to infer from the PNR histogram: the mean integrated flux denoted as $Phe\ rate$ of the Poisson distribution, the mean and sigma of the Gaussian read noise, respectively denoted as $RN\ mu$ and $RN\ sigma$, the mean of the Gamma distribution, denoted as $avrg\ avlch\ gain$, and the $ENF$. The sigma of the Gamma distribution is constrained, once the values of $avrg\ avlch\ gain$ and $ENF$ are fixed, by the relation Equ.~\ref{eq:enf}. The parameter $avrg\ avlch\ gain$ is indistinguishable from the charge gain given only the information contained in the PNR histogram due to the degeneracy of these two quantities in the conversion gain. Therefore, it is known to within the precision of the charge gain which is assumed to be the same for every pixel and fixed at 3.6 e-/ADU. This value comes from the measurement of the voltage gain of the detector and the pixel node capacitance provided by Leonardo. The inferred value of $avrg\ avlch\ gain$ scales inversely proportional to the fixed value of the charge gain, so that the derived value of the conversion gain, which is the ratio of these two quantities, remains unchanged. 

\subsection{Test implementation}

For this test, the detector was configured at a bias voltage of 13V and read out at the nominal speed (1.45 FPS). To further reduce the read noise through multisampling and averaging according to a 25-Fowler sampling scheme, each individual exposure consisted of: 25 read frames taken at the beginning of the exposure, followed by $\sim$10 s of integration time, followed by 25 more read frames. A total of 5040 individual exposures were taken, each lasting $\sim$45 s, resulting in a total test duration of 3.5 days. This configuration was primarily motivated by the results shown Fig.\ref{fig:rn_fwlr_sampling} to achieve a sufficiently low read noise and an adequate number of samples within a reasonable time. Finally, the VOA was synchronized with the data acquisition: it was closed and blocked the light flux during the 25 read frames and opened during the 10s integration time, to avoid receiving photons during the read frames, which would have increased the effective read noise, even at low flux. The light flux was adjusted to receive on average $\sim$1 phe- per exposure, or PNR samples. The glow was sufficiently reduced by a factor of 14 as described Sec.~\ref{sec:perf:red_ron} to be considered negligible.

The statistical model described above was implemented using the Python packages PyMC\cite{pymc2023} and JAX\cite{jax2018github}. The discretization by the ADC was neglected. % - although it would also be possible to add it to the PyMC model thanks to the “round" function and the automatic likelihood derivation offered by PyMC.
The number of components in the mixture distribution is limited to 15. The PNR parameters are inferred from the PNR histogram using the MAP (maximum a posteriori) solver of PyMC, and their uncertainties are estimated using 100 bootstrap resampling of the PNR histogram. The prior distributions adopted for each PNR parameter are summarized in Tab.~\ref{table:pnrpriors}. After aggressive filtering of the pixel and PNR samples, 50 “best" pixels were selected out of $\sim$7300 analyzed, each with a PNR histogram made up of $\sim$3000 PNR samples. The average read noise expected for this subset of pixels is $\sim$2.5 ADU.

\begin{table}[htbp]
\centering
\caption{Prior distributions, along with their adopted parameters, used for each of PNR parameter inferred from the PNR histogram. The lower and upper limits given for the Uniform distribution are noted respectively ``lwr'' ``upr''.\newline}
\label{table:pnrpriors}
\begin{tabular}{lc}
\textbf{PNR parameters}                                            & \textbf{Prior distributions}                 \\ \hline
\textbf{Phe rate {[}$\mathbf{\textbf{ph.e}^{-}/\textbf{sec}}${]}} & Gamma(mu=+1.0, sigma=+1.0)                   \\
\textbf{RN mu {[}ADU{]}}                                           & Uniform(lwr=-10.0, upr=+10.0)                \\
\textbf{RN sigma {[}ADU{]}}                                          & HalfNormal(sigma=+1.0)                       \\
\textbf{avrg avlch gain {[}1{]}}                                   & Uniform(lwr=+10.0, upr=+60.0)                \\
\textbf{log10(ENF-1) {[}1{]}}                                      & Uniform(lwr=log10(+1.0e-5), upr=log10(+1.0)) \\ \hline
\end{tabular}
\end{table}

\subsection{First results}

\begin{figure}[htbp]
  \begin{subfigure}{0.49\columnwidth}
    \includegraphics[width=\linewidth]{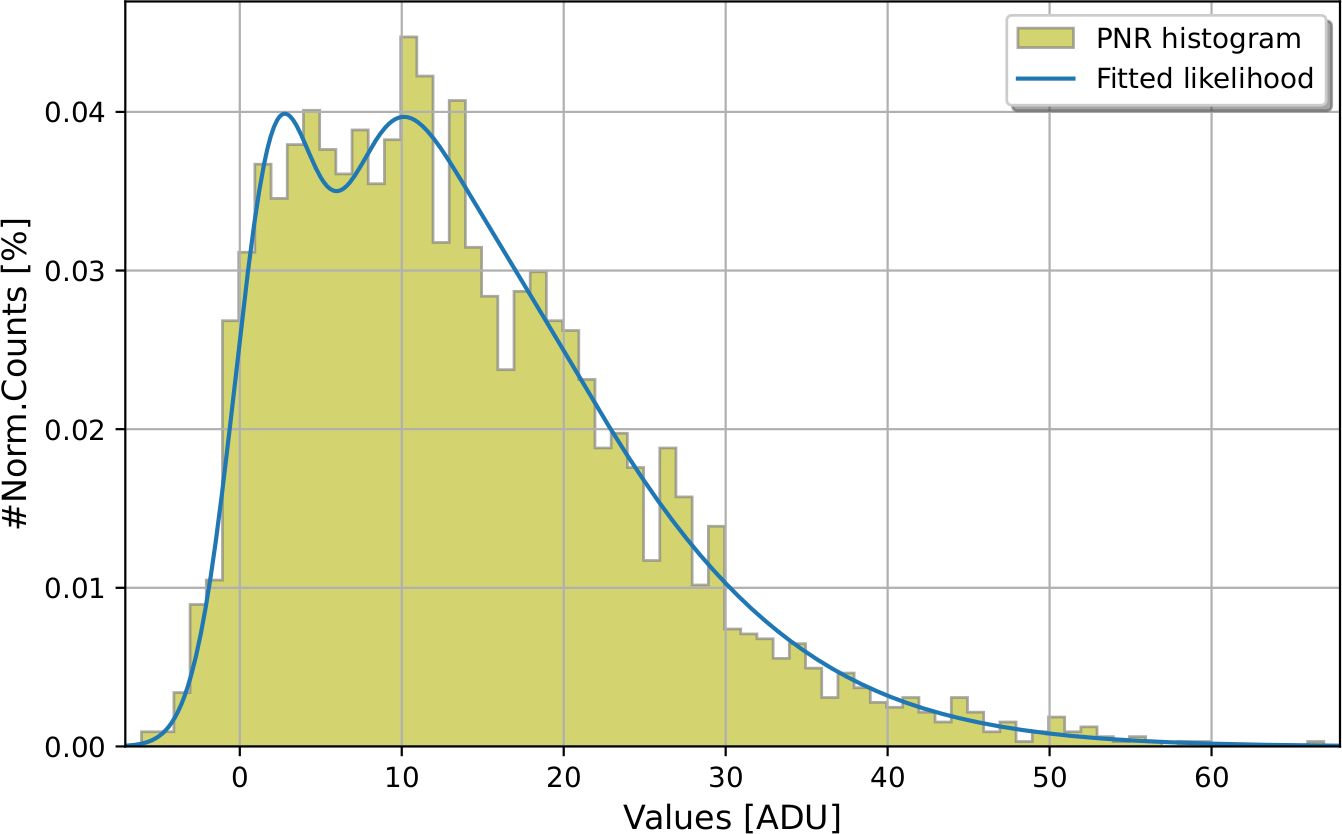}
    \caption{Original (i.e., non-resampled with bootstrap) PNR histogram in yellow bars obtained for one selected pixel. It contains 3242 samples. The system parameters inferred from this PNR histogram using the MAD solver are: phe rate = 1.56 phe$^{-}$/exp, RN mu = 2.0 ADU, RN sigma = 2.5 ADU, average avalanche gain = 27.7, ENF = 1.20, resulting in a ``true'' conversion gain of 0.13 e-/ADU, and an effective RN of 0.32 e-. The likelihood function evaluated with the inferred parameters above is shown in blue overlapped on the histogram.\newline\newline}\label{fig:real_pnr_hist}
  \end{subfigure}\hfill
  \begin{subfigure}{0.49\columnwidth}
    \includegraphics[width=\linewidth]{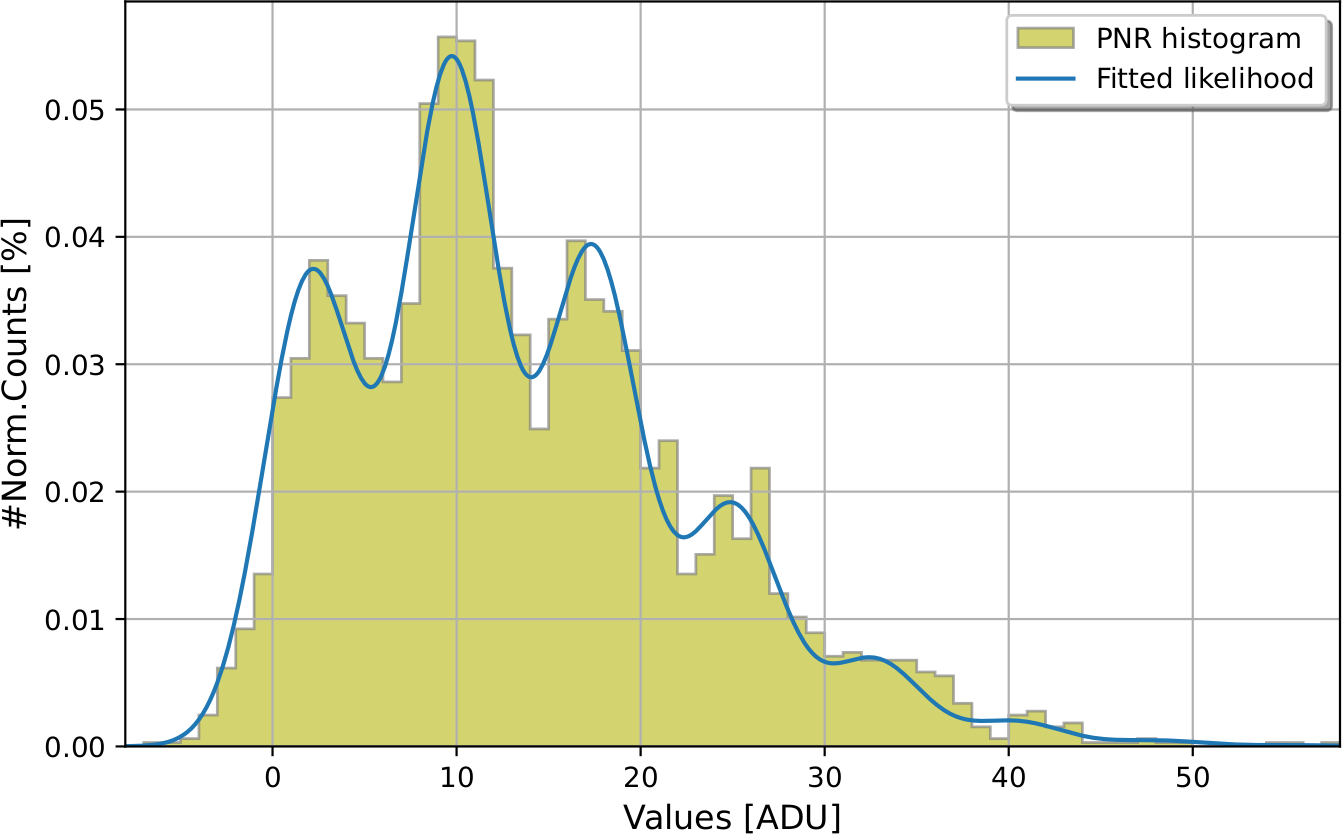}
    \caption{Simulated PNR histogram in yellow bars using input parameters equal to the parameters inferred from the ``real'' PNR histogram shown on the left Fig.~\ref{fig:real_pnr_hist}, except that the ENF is here fixed to 0 in order to highlight the difference with vs. without ENF. It contains 3250 samples. The system parameters inferred from this simulated PNR histogram using the MAD solver are: phe rate = 1.45 phe$^{-}$/exp, RN mu = 2.1 ADU, RN sigma = 2.5 ADU, average avalanche gain = 27.6, ENF = 1.000, resulting in a “true" conversion gain of 0.13 e-/ADU, and an effective RN of 0.33 e-. The likelihood function evaluated with the inferred parameters above is shown in blue overlapped on the histogram.}\label{fig:sim_pnr_hist}          
  \end{subfigure}\newline
  \caption{Comparison of PNR histograms from real and simulated data.}
  \label{fig:pnr_hist}
\end{figure}

\begin{figure}[htbp]
  \centering
  \includegraphics[width=1.\linewidth]{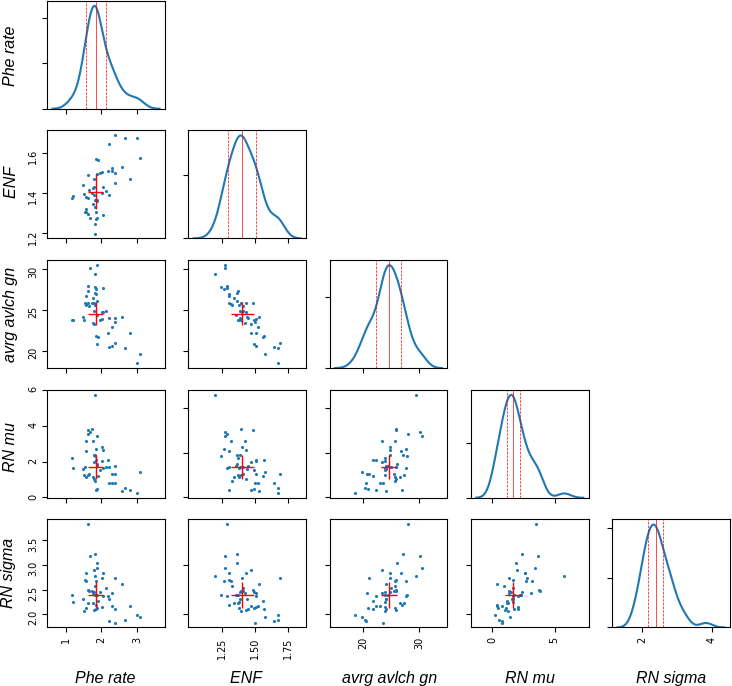}
  \vspace{\baselineskip}
  \caption{Pair plot showing the scatter plots for each pair of PNR parameters inferred from the PNR histograms of the 50 selected pixels in blue points, along with the corresponding KDE histograms of their pixel distributions. The median of the pixel distributions is indicated by a solid red vertical line, while the two dotted red vertical lines positioned on either side, are evenly spaced from the median by the same distance equal to the MAD of the distribution. The error bars on the scatter plots indicates the median and MAD values of the pixel distributions.}
  \label{fig:pnr_pair_plot}
\end{figure}

An example of a PNR histogram obtained for one of the 50 selected pixels is shown on the left Fig.~\ref{fig:real_pnr_hist}, see description in the legend. The estimated ENF is 1.20. For comparison, it is also shown on the right Fig.~\ref{fig:sim_pnr_hist}  what the PNR histogram would look like for the same PNR parameters, except for the ENF fixed at 1. One can appreciate how the ENF affects the shape of the PNR histogram, smoothing the PNR histogram and making it more difficult to discern the individual peaks, thus hindering the photon number resolving capability of the detector. The pixel distributions of the inferred PNR parameters for the 50 selected pixels are shown Fig.~\ref{fig:pnr_pair_plot}, and the statistical summary is given in Tab.~\ref{table:pnr}. The mean pixel STD is comparable to or greater than the MAD value for all PNR parameters, suggesting that the width of the pixel distributions is dominated by the statistical uncertainty obtained via bootstrap method, due to the limited number of samples of the PNR histograms.

The validity of the analysis method was evaluated using simulated data taking the median values given in Tab.~\ref{table:pnr} for the PNR parameters used in the simulations. The results showed good compatibility with the mean values and associated statistical uncertainties returned by the bootstrap method. The estimated systematic error for each PNR parameter is summarized in Tab.~\ref{table:pnrbias}. It is on the order of 5\% (except for $RN\ mu$), and is on average a factor of 2 smaller than the statistical error. This systematic error may arise from degeneracy between parameters, especially as the ENF increases, or from a misspecification of the model, such as the choice of the maximum number of components in the mixture distribution - e.g., if there are some samples, even just a few, in the measured PNR histogram to fit that are located in the tail of the “true" distribution and not covered by any component included in the mixture.

Note that any possible distortion of the PNR histogram due to residual glow (generated and accumulated during the detector readout within the two Fowler frame subsets at the beginning and end of each PNR exposure) or due to tunneling dark current (accumulated both during the detector readout and also within the integration window) at BV=13V has not been taken into account in this study. This would need to be investigated by simulation to verify and confirm that these two extra sources of signal do not introduce bias into the reported results, especially since these two components are generated according to a Poisson distribution, which could introduce skewness into the peaks of the mixture distribution. However, two elements suggests they have a negligible impact. Firstly, unlike ENF, which causes increasing broadening of the peaks with the number of photoelectrons received per peak, since each absorbed photoelectron experiences the avalanche process independently, the accumulation of glow or DC will result in identical broadening for every peak. Consequently, any broadening of the peaks will be preferably attributed to an increase in $RN\ sigma$, and not to the ENF. Secondly, a control dataset was acquired following the same data acquisition configuration but under dark conditions and for 100 samples prior to the PNR test with illumination of the detector. For each of the 50 selected pixel, the “dark" PNR histogram constructed from the control dataset directly corresponds to the distribution of the read noise (in ADU) reduced by multisampling + averaging. Their analysis shows distributions without notable skewness and with a STD comparable to the $RN\ sigma$ measured from the PNR histograms of the selected pixels and reported below. Note also that glow and dark current do not necessarily share the same conversion gain as photoelectron signals, and that it might be useful to measure it using PTC.

\begin{table}[htbp]
\centering
\caption{Median value and scaled MAD (median absolute deviation multiplied by $\sim1.48$ to be equivalent to the STD of the distribution assumed to be Gaussian) of the pixel distribution obtained for each of the five PNR parameters inferred from analysing the PNR histograms associated with the 50 selected pixels. As a comparison, the mean pixel STD is also indicated. It is the average over the pixels of their parameter uncertainties estimated using the boostrap method applied to the PNR histograms.\newline}
\label{table:pnr}
\begin{tabular}{@{}lcc@{}}
\textbf{PNR parameters}                                            & \textbf{Pixel median $\pm$ MAD} & \textbf{Mean pixel STD} \\ \midrule
\textbf{Phe rate {[}$\mathbf{\textbf{ph.e}^{-}/\textbf{exp}}${]}} & $1.84 \pm 0.21$                 & $0.28$                  \\
\textbf{RN mu {[}ADU{]}}                                           & $1.70 \pm 0.66$                 & $0.54$                  \\
\textbf{RN sigma {[}ADU{]}}                                          & $2.40 \pm 0.26$                 & $0.21$                  \\
\textbf{avrg avlch gain {[}1{]}}                                   & $24.6 \pm 1.4$                  & $2.3$                   \\
\textbf{ENF {[}1{]}}                                               & $1.41 \pm 0.087$                & $0.10$                  \\ \bottomrule
\end{tabular}
\end{table}

\begin{table}[htbp]
\centering
\caption{Systematic uncertainty for each of the five PNR parameters and estimated by applying the analysis method on randomly-generated simulated data, with a total of 100 realizations. One simulated dataset correspond to a PNR histogram of 3000 samples. The PNR parameter values adopted for the simulation are the same as the pixel median values given in Tab.~\ref{table:pnr}.\newline}
\label{table:pnrbias}
\begin{tabular}{lc}
\textbf{PNR parameters}                                            & \textbf{Systematic uncertainty} \\ \hline
\textbf{Phe rate {[}$\mathbf{\textbf{ph.e}^{-}/\textbf{sec}}${]}} & -5.5\%                          \\
\textbf{RN mu {[}ADU{]}}                                           & +10\%                           \\
\textbf{RN sigma {[}ADU{]}}                                          & +2.1\%                          \\
\textbf{avrg avlch gain {[}1{]}}                                   & +6.4\%                          \\
\textbf{ENF {[}1{]}}                                               & -4.2\%                          \\ \hline
\end{tabular}
\end{table}

In conclusion, it is interesting that the method developed in this study, and applied for the first time for a few pixels on data taken with the 1kx1k science-grade sensor, returns median values for $Phe\ rate$ (1.8$\pm$0.2 phe-/exp) and $RN\ sigma$ (2.40$\pm$0.3 ADU) consistent with what is expected given the test configuration. The median value found for the ENF is 1.4$\pm$0.09, which in principle does not allow for photon number resolving at the considered bias voltage of 13V. Such an ENF value results in a broadening of the peak associated with single-photon events of the PNR histogram equivalent to an effective read noise of \mbox{$\sqrt{RN^{2} + (ENF-1)^{2}}=\sqrt{(0.3)^{2} + (0.4)^{2}}=0.5$ e-}. It must be noted that this is still a preliminary estimation that should be taken with caution. The inferred median value for $avrg\ avlch\ gn$ is 24.6$\pm$1.4. As a result, the predicted value of the conversion gain measured by PTC is $3.60/(24.6\times1.4)=0.10\pm0.009$ e-/ADU, which is consistent with the actual PTC measurement ($\sim$0.09 e-/ADU) for 13V of bias, reported Tab.~\Ref{table:glob_perfs}, given the uncertainties. This also means that the conversion gain measured by PTC may be underestimated by $\sim$40\%, and its ``true'' (ENF-corrected) value should be $0.14\pm0.008$ e-/ADU. The estimated effective read noise is $2.4\times0.14=0.34\pm0.04$ e-, given 13V of bias, combined with 25-Fowler multisampling. Therefore, the method proves to be a valuable calibration tool at the pixel level, despite a limited number of PNR samples per pixel, and an ENF value for which the predictive power of the PNR histogram might be reduced. Finally, the results also support the validity of the Gamma Law for representing the distribution of APD gains.

\section{On-sky demonstration}

On-sky testing was conducted in two runs of three nights each in August 2023. The cryostat was reconfigured into camera mode. The integrating sphere was removed, a system of four filters that can be manually switched was added, and the top cover was replaced with the field lens. The four available filters consists of two broadband filters, namely an H-band and a J-band, a narrow CH4-band filter centered at $\lambda$=1.6$\mu$m with approximately 100nm bandpass, and an aluminum blocking filter. The camera was installed at the Cassegrain focus at the back of the UH2.2m telescope at Mauna Kea. A new mask was placed above the detector, providing a larger central square area exposed to illumination, namely 65\% compared to only 1\% with the mask used during lab testing. The edges of the mask cover a band approximately 100 pixels wide on each side of the detector. This detector region shielded from light serves as a control area and can be useful for monitoring dark current and persistence. The cryostat and detector were operated in exactly the same way as during lab tests and cooled to 50K.

The purpose of the on-sky testing was to demonstrate the application and photometric capabilities of the detector for astronomical observations in simple CDS and UTR-sampling modes. Primarily, two types of targets were selected and observed:
\begin{enumerate}
    \item Visually complex and structured objects to demonstrate the detector’s ability to capture “photogenic" images rich in details.
    \item Variable objects characterized by amplitude variations from a few percent to tens of percent, over timescales ranging from minutes to a few hours, aiming to demonstrate the detector’s ability to monitor flux variations over time.
\end{enumerate}

Additionally, other auxiliary data were acquired for target image reduction, such as flat fields, sky background, dark integrations, as well as calibration data, e.g., for measuring the detector’s linearity or for PTC measurement via temporal averaging. In total, nearly 50GB of data were collected and are still being analyzed.

\begin{figure}[htbp]
  \begin{subfigure}{0.49\columnwidth}
    \includegraphics[width=\linewidth]{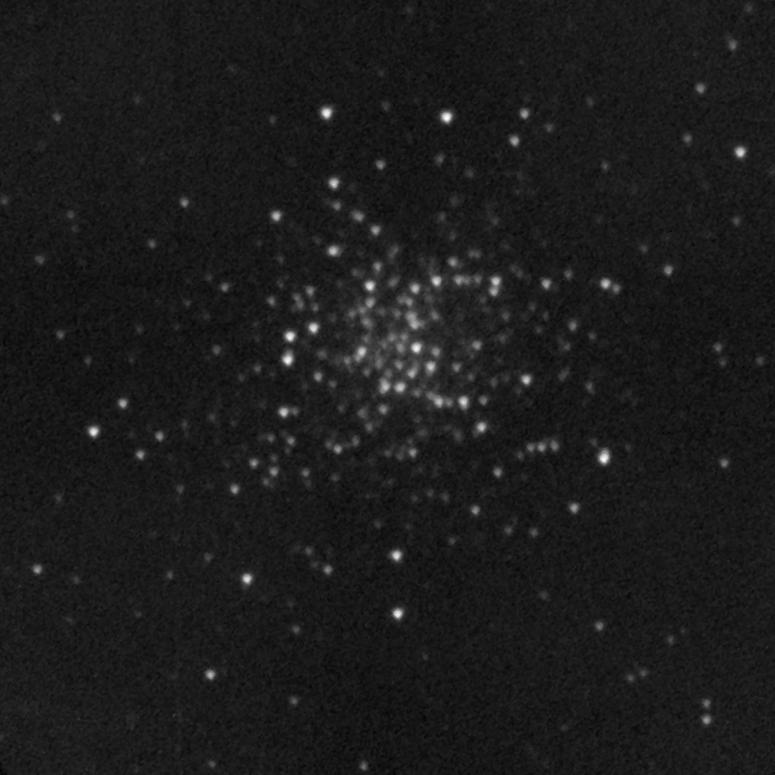}
    \caption{Image taken with our LmAPD array on the UH2.2m telescope using the CH4-band filter.}\label{fig:palomar2_lmapd}
  \end{subfigure}\hfill
  \begin{subfigure}{0.49\columnwidth}
    \includegraphics[width=\linewidth]{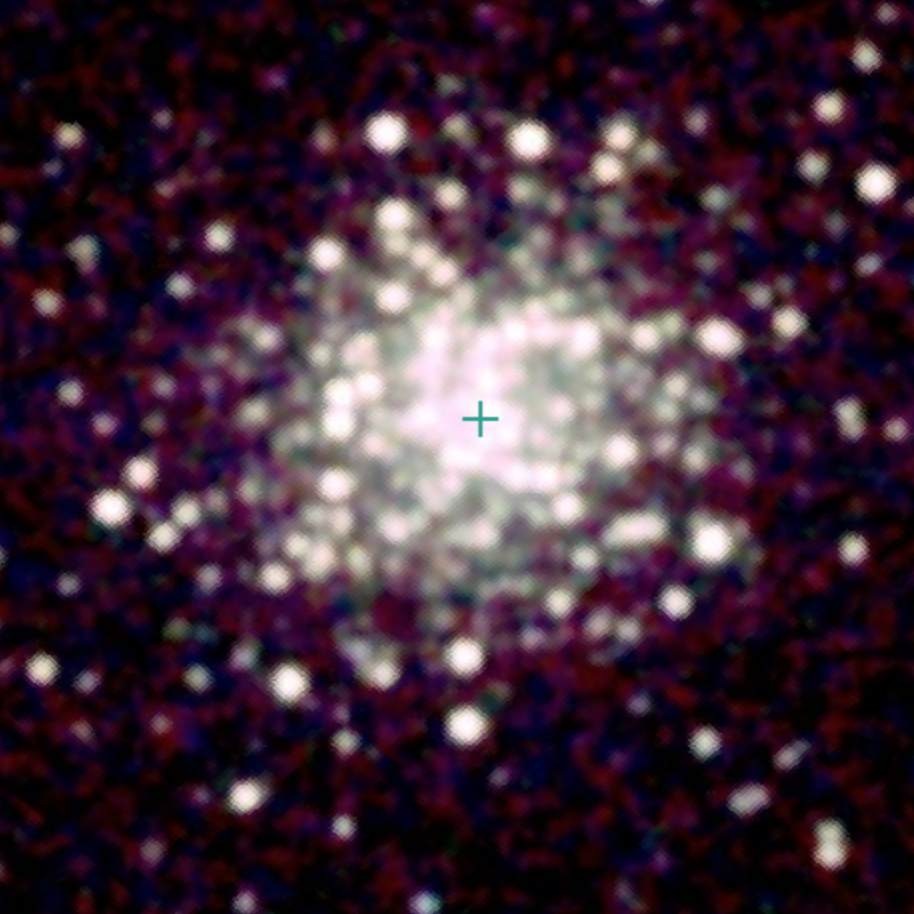}
    \caption{Image taken from the 2MASS survey with a 1.3 m telescope using an H-band filter.}\label{fig:palomar2_2mass}
  \end{subfigure}\newline
  \caption{Image of the Palomar 2 object taken with our LmAPD array compared to the image of this same object observed by the 2MASS survey.}
  \label{fig:palomar2}
\end{figure}

\begin{figure}[htbp]
  \begin{subfigure}{0.49\columnwidth}
    \includegraphics[width=\linewidth]{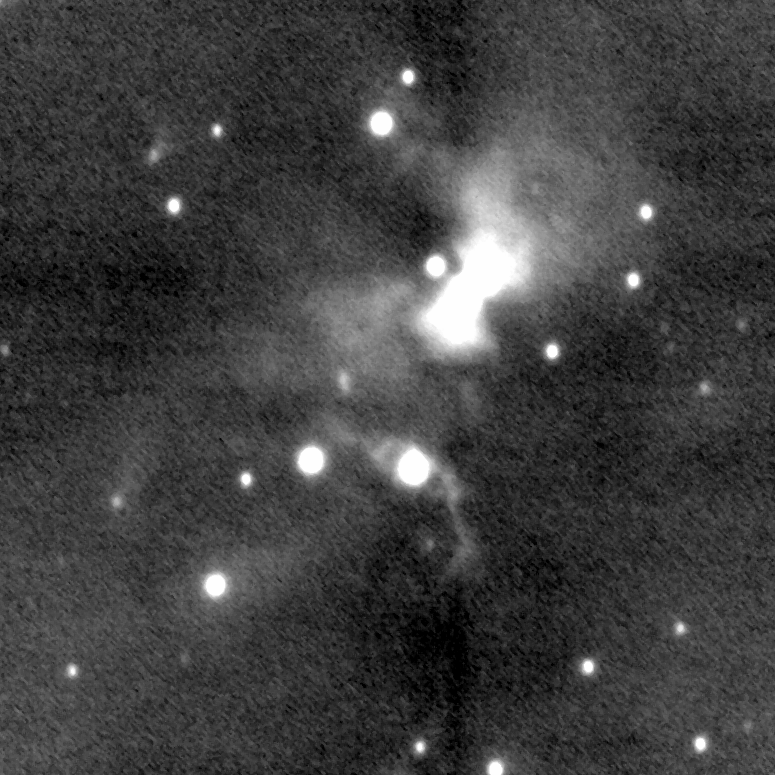}
    \caption{Image taken with our LmAPD array on the UH2.2m telescope using the CH4 band filter.}\label{fig:serpens_lmapd}
  \end{subfigure}\hfill
  \begin{subfigure}{0.475\columnwidth}
    \includegraphics[width=\linewidth]{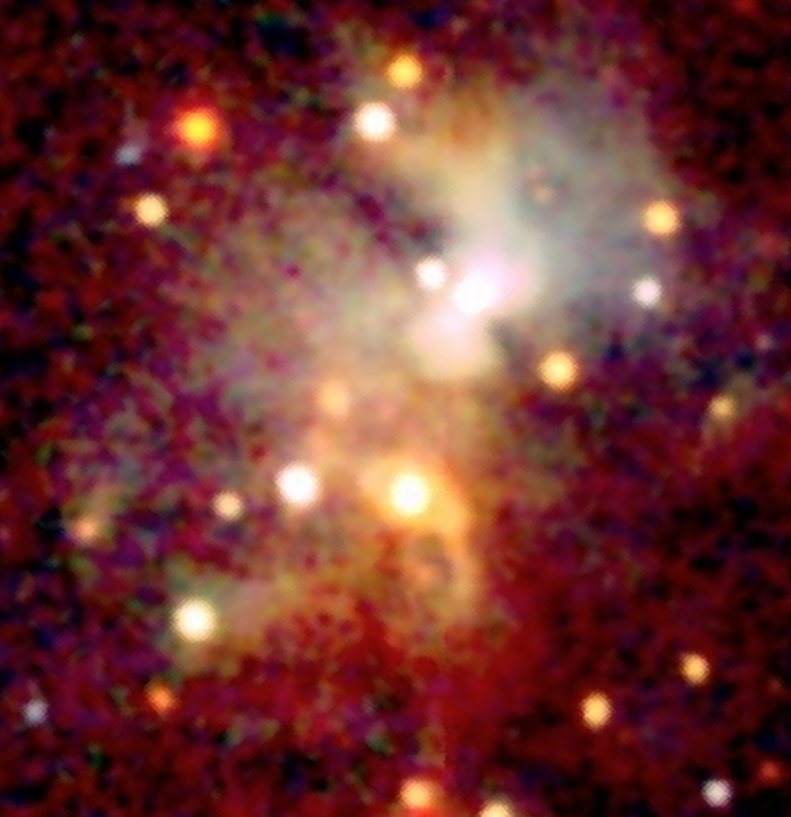}
    \caption{Image taken from the 2MASS survey with a 1.3 m telescope using an H-band filter.}\label{fig:serpens_2mass}
  \end{subfigure}\newline
  \caption{Image of the Serpens Main object taken with our LmAPD array compared to the image of this same object observed by the 2MASS survey.}
  \label{fig:serpens}
\end{figure}

Two examples of “photogenic" images taken with the LmAPD arrays are shown in Fig.~\ref{fig:palomar2} and Fig.~\ref{fig:serpens}, taken using the CH4-band filter to reduce sky background contamination. These images are compared with the colorized versions of the same objects taken during the 2MASS survey\cite{Skrutskie_2006} (Two Micron All-Sky Survey). The 2MASS survey mapped the entire sky in three infrared wavelength bands (J-, H-, and Ks-band) between 1997 and 2001 using two 1.3m telescopes, one located in the northern hemisphere and the other in the southern hemisphere. Fig.~\ref{fig:palomar2_lmapd} shows an image of Palomar 2, which is a faint globular cluster located in the constellation Auriga in our galactic halo. The resolution of the image offers a clear distinction of the cluster’s member stars. Fig.~\ref{fig:serpens_lmapd} shows Serpens Main, a dense molecular region rich in gas and dust, very active in star formation, and located in our galactic disk. Many features and details of the gas cloud are recognizable in this image. The first image was obtained by stacking the equivalent of 126 frames with 1.4s exposure time, and the second image by stacking the equivalent of 240 frames with the same exposure time.

\newpage
\section{Conclusion}

\subsection{Other detector properties}
\label{sec:concl:other_effects}

\begin{figure}[htbp]
  \centering
  \includegraphics[scale=0.65]{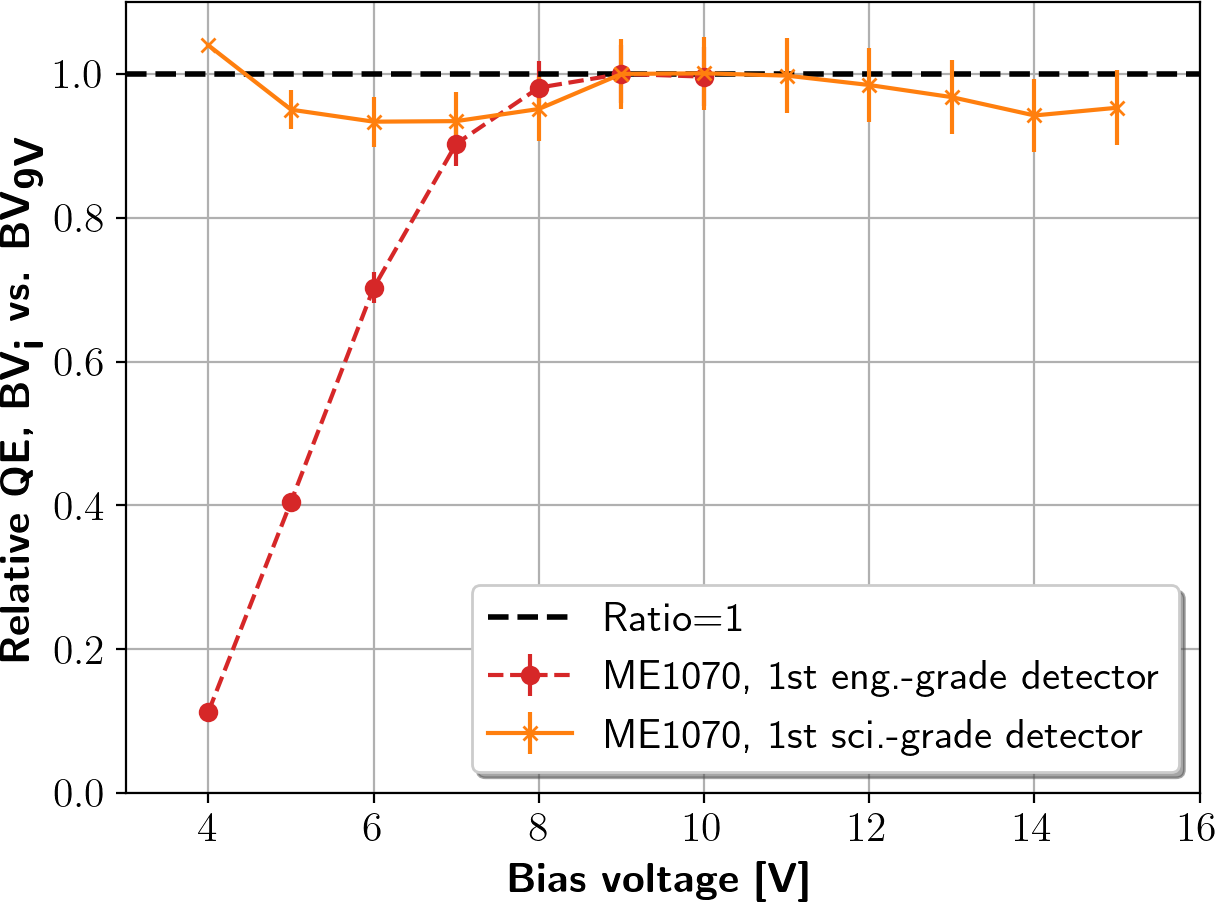}
  \vspace{\baselineskip}
  \caption{Relative QE vs. bias voltage. The curves shows the increase of the signal ratio taking bias voltage=4V as reference, and corrected for the measured conversion gain ratio. Due to a contamination issue in the engineering-grade devices, its QE drops by nearly a factor 9 and quasi-linearly from bias voltage=8 to 4V. Beyond bias voltage=8V, the QE reaches a plateau. In contrast, the science-grade detector shows a uniform quantum efficiency at all bias voltages.}
  \label{fig:qe_ggn_consistency}
\end{figure}

Several other properties of the detector were investigated:

\begin{itemize}
    \item Quantum efficiency - The \textgreater80\% reduction in relative quantum efficiency evident in the engineering-grade sensors at low bias voltages has been eliminated in the science grade devices as can be seen Fig.~\ref{fig:qe_ggn_consistency}, which demonstrate uniform quantum efficiency at all bias voltages.
    \item Uniformity – Under flat-field illumination, we measure a detector uniformity of 18\%, at T=50K. This number is defined as the ratio of the sigma value over the peak value of the pixel distribution obtained from a detector sub-region uniformly illuminated. This is consistent with manufacturer measurements. This non-uniformity is higher than expected, and improves with temperature. It is being investigated with Leonardo, and is suspected to come from a variability of QE between pixels.
    \item Persistence - After exposing the central region of the detector to continuous illumination in several tests, it was observed that this area retained memory of previous light exposures. This results in a strong persistence effect and an increased intrinsic dark current in that specific region of the detector. At high BVs (\textgreater10.0V), persistence could cause a more than 20-fold increase in dark current, primarily due to tunneling current.  This is consistent with the idea that tunneling and persistence are both due to traps in the material. Persistence and mitigation effects are still investigated. The only effective way found so far to clean up the persistence while keeping the detector at T=50K, albeit without being able to return to its initial state free of any residual persistence, is to continuously cycle the detector at very high bias voltage (\textgreater14V), repeating the same sequence of reset + read frames. The solution to completely remove persistence is to perform an annealing operation of the detector at room temperature. 
    \item Random telegraph signal - we confirmed several pixels exhibit what appears to be random telegraph signal (RTS), a known effect caused by spontaneous change of the pixel value between two levels, believed to be caused by trapping and de-trapping electrons.
\end{itemize}

\subsection{Summary of the results}
\label{sec:concl:sum}

\begin{itemize}
    \item The science-grade devices have solved many of the issues inherent in the engineering grade device.  Their performance comes very close to, and in some cases exceeds, the initial goals of this investigation.  This includes measurements of readout noise and dark current.
    \item The graded bandgap design has been a success, pushing out the onset of tunneling current to higher bias voltages.
    \item The science-grade devices show marked improvement in the number of pixels exhibiting tunneling current. At the optimal operating point, only a few percent of pixels show anomalously high dark current.
    \item Using frame averaging, an effective read noise as low as $\sim$0.3 e- has been demonstrated.
    \item A first attempt to demonstrate the photon number resolving capability of the detector was conducted. The developed analysis method proved to be a powerful calibration tool for characterizing the detector a the pixel level based on their PNR histograms. A preliminary estimation of the ENF$\sim$1.4 was obtained for a bias of 13V. Such an ENF value prevents the distinction of individual photo events
    \item On-sky testing of the detector was conducted under conditions similar to those it will encounter in an astronomical environment, demonstrating its performance for astronomical observations.
    \item The science-grade sensors are now at TRL-4, having been operated successfully in a laboratory environment for $\sim$2 years.
\end{itemize}

\subsection{Path forward}
\label{sec:concl:next_steps}

A more in-depth characterization of the persistence effect observed in our data is needed at different temperatures and bias voltages. Persistence and non-uniformity are suspected to be related and possibly have a common origin. Investigation has already started in collaboration with Leonardo who are conducting new series of experiments and iterations of the bandgap structure. In parallel, work is in progress to establish an accurate physical model of the electric field in the photodiode structure. All these ongoing developments will contribute to gain a more detailed understanding of the photodiode, enabling further optimization and performance improvements. They will help continue pushing the onset of tunneling current to higher bias voltages, minimize the occurrence of outliers, maintain high APD gains to further reduce effective read noise, and limit the ENF.

So far, most of the detector's testing has been carried out at T=50K. It would be valuable to extend our knowledge of the detector's properties to other temperatures, in particular to compare the results with theoretical predictions, and revise our model if need be. A radiation testing campaign will be conducted on the science-grade detectors to advance their qualification for space applications. A complete characterization of the detectors will be performed before and after irradiation to investigate the effects of radiation on the device properties and assess any degradation in performance. It would be interesting to replicate the PNR tests and ENF measurements for more pixels, with more samples in the PNR histograms, and perhaps at different photon rates and other bias voltages, in order to verify the preliminary results presented in this paper. For example, the decreasing trend visible on the orange curve of the relative QE in Fig.~\ref{fig:qe_ggn_consistency} starting from a bias \textgreater11V may partly be due to an increase of the ENF at high bias voltages. Potential experimental systematics that may affect the ENF measurements must be carefully examined. Detector sub-windowing could prove useful to speed up data collection.

A new, larger 2kx2k version of the LmAPD detector is currently under development in collaboration with Leonardo. It holds the potential to offer exceptionally high-performance drop-in detectors for existing on-ground instruments already equipped with H2RGs, and is anticipated to be highly beneficial for high-resolution spectroscopy applications.

\acknowledgments % equivalent to \section*{ACKNOWLEDGMENTS}
 
The authors would like to thank Dr. Gert Finger of ESO for his helpful comments on our efforts. Development of the sensors and their characterization at UH is sponsored by NASA SAT award \#18-SAT18-0028. 

% References
\bibliography{mybiblio} % bibliography data in report.bib
\bibliographystyle{spiebib} % makes bibtex use spiebib.bst

\end{document}